%% file: conference_101719.tex
\documentclass[conference]{IEEEtran}
\IEEEoverridecommandlockouts
\usepackage{cite}
\usepackage{amsmath,amssymb,amsfonts}
\usepackage{algorithmic}
\usepackage{graphicx}
\usepackage{textcomp}
\usepackage{xcolor}
\usepackage{multirow}
\usepackage{makecell}
\usepackage{graphicx, subcaption}
\def\BibTeX{{\rm B\kern-.05em{\sc i\kern-.025em b}\kern-.08em
    T\kern-.1667em\lower.7ex\hbox{E}\kern-.125emX}}
\begin{document}
\date{}
\def\NAME{\textsc{SysFlow}}

\makeatletter
\newcommand{\linebreakand}{%
  \end{@IEEEauthorhalign}
  \hfill\mbox{}\par
  \mbox{}\hfill\begin{@IEEEauthorhalign}
}
\makeatother


\title{\NAME{}: Efficient Execution Platform for IoT Devices\\
}
 \author{IEEE INFOCOM 2024, Paper ID: \#0000}
 \author{\IEEEauthorblockN{Jun Lu}
 \IEEEauthorblockA{\textit{Central South University} \\
 jun.lu@csu.edu.cn}
 \and
 \IEEEauthorblockN{Zhenya Ma}
 \IEEEauthorblockA{\textit{Tsinghua University} \\
 mzy23@mails.tsinghua.edu.cn}
 \and
 \IEEEauthorblockN{Yinggang Gao}
 \IEEEauthorblockA{\textit{Central South University} \\
 yinggang\_gao@163.com}
 \linebreakand
 \IEEEauthorblockN{Ju Ren}
 \IEEEauthorblockA{\textit{Tsinghua University} \\
 renju@tsinghua.edu.cn}
 \and
 \IEEEauthorblockN{Yaoxue Zhang}
 \IEEEauthorblockA{\textit{Tsinghua University} \\
 zhangyx@tsinghua.edu.cn}
 }


\maketitle

 \input{0_abstract.tex}
 \input{1_introduction.tex}
 \input{2_background.tex}
 \input{3_design.tex}

 \input{4_implementation.tex}
 \input{5_evaluation.tex}
 \input{6_relatedwork.tex}
 \input{7_conclusion.tex}




\bibliographystyle{IEEEtran}
\bibliography{references}

\end{document}

%% file: 0_abstract.tex
\begin{abstract}
Traditional executable delivery models, such as RPM, pose challenges for IoT devices with limited storage capacity, requiring the download of complete suites of executables and their dependencies. Network solutions like NFS attempt to address this by enabling dynamic loading during runtime. However, they are designed for data files and encounter high network and disk IO overhead when delivering executables with irregular access patterns.

In response to these challenges, this paper introduces \NAME{}, a lightweight network-based executable delivery system designed to operate efficiently in IoT environments. \NAME{} delivers parts of executables on-demand, resembling a stream. It strategically redirects local disk IO for accessing executables to the server through optimized network IO.
To optimize the page cache hit rates of client applications, \NAME{} incorporates an action-based prefetching mechanism on the server side. This mechanism asynchronously pre-reads block sequences with significant differences into memory, based on normalized variance, effectively concealing disk IO latency in advance.
Experimental results demonstrate that \NAME{} achieves a latency reduction of 45.1\% to 75.8\% compared to native Linux filesystems using SD cards. In wired environments, \NAME{}'s latency is up to 67.7\% lower than NFS. In wireless scenarios, \NAME{}'s performance is only 22.9\% worse than Linux filesystems, making it comparable with Linux and outperforming NFS by up to 60.7\%. While \NAME{}'s power consumption may be at most 6.7\% higher than NFS, it offers energy savings due to lower processing time.

\end{abstract}

%% file: 1_introduction.tex
\section{Introduction}
The traditional executable delivery model is to download the complete executable binaries and dependencies like required dynamic libraries to local storage.
This method is commonly used by popular package management systems such as RPM and DEB~\cite{ewing1996rpm, blackman2000debian}.
As software functions become increasingly abundant, the software size also increases, leading to software bloat. 
Research has shown that only 20\% of executables are used, resulting in excessive storage occupied by unused code \cite{holzmann2015code, quach2017multi, muller2020exploration, soto2021comprehensive}.
 IoT devices, such as smart wearable devices, often suffer from a lack of sufficient storage, further exacerbated by the increasing software size. 
The extension of storage for these devices is constrained by various factors, including the physical size of the device, the difficulty of replacing storage such as Flash and eMMC, and the cost of upgrading to higher-capacity options.

In recent years, researchers have proposed a variety of software debloating solutions, which eliminate and remove unnecessary functionalities from the programs. Typical software debloating techniques~\cite{qian2019razor, quach2018debloating, soto2022coverage, qian2020slimium, zhang2022one} include both static methods like binary scanning, dynamic analysis like runtime function tracing and machine-learning-based predictions~\cite{porter2020blankit}. 
Various software debloating methods have the potential to significantly reduce executable size and minimize the attack surface at the software level.
However, some incorrect predictions and trims may cause application crashes in corner cases.

Meanwhile, network-based solutions\cite{nfs, efs, ghemawat2003google, weil2006ceph, anderson2020assise} enable clients to dynamically load required parts of executable files from the server on demand during runtime.
However, these solutions rely on the server's file system, which is optimized for data files with sequential and stride access patterns rather than executables with unpredictable behavior. 
For example, although NFS~\cite{nfs} can achieve extreme diskless by saving the entire software on the server, its inefficient prefetching mechanism results in a more significant number of network round trips when the client executes files with jump instructions.

In this paper, we present \NAME{}, a lightweight online executable delivery system, which adheres to the client-server model. 
The server stores entire applications (including executables and their dynamic libraries) in disks or memory. The client will load \textit{.text} section of ELF to directly run the applications, avoiding unused code blocks wasting limited storage space. 
When the client runs an application, \NAME{} on the client side intercepts the disk IO, identifies the blocks required by the application, and redirects them to the server.
Once the server receives the requests, \NAME{} predicts the blocks that may be needed soon, retrieves requested and predicted blocks from the disk or memory, and sends them to the client via the network.

However, implementing \NAME{} faces several challenges. 
Firstly, it is difficult for \NAME{} to anticipate when the application  initiates disk IO. It consequently causes delay in resource allocation, like establishing and releasing network connections and allocating memory for the code block stream. 
Secondly, unlike data files that exhibit sequential or stride access patterns, executables have much more unpredictable patterns due to complex jump/call instructions. 
Even with POSIX \textit{fadvise} syscall, it is difficult for applications to provide accurate hints about access patterns of executables.
Though fine-tuned machine learning models can provide accurate predictions\cite{al2020effectively}, the millisecond-level inference and substantially high training overhead render this approach unacceptable.
Lastly, edge servers with limited memory cannot always cache executables in memory to keep them warm, which causes excessive disk IO once they are not in memory.
Although page cache management strategies such as LRU may help alleviate this situation, they can cause cache thrashing when working on IoT scenarios with huge working sets.

Initially, we leverage traditional asynchronous methods to optimize network IO, preventing resource preparation blocking. This involves deploying asynchronous threads like the Networker for maintaining persistent connections and the Allocator for managing page pools.
Observing block access patterns, we identify that application tasks often exhibit continuity in code block stream requests within a specific timeframe. These tasks consistently and repeatedly load fixed code blocks, showcasing stability and repeatability. Each task is abstracted into an action, containing the historical block stream sequence of the executable for matching. On the server side, \NAME{} attempts to match the client's block request with existing actions. Upon success, \NAME{} uses the action to predict and advance-push accurate future block streams. In the absence of a match, \NAME{} generates a new action for the requested block stream.
Lastly, to prevent cache occupation by unused blocks, server-side \NAME{} reads necessary code blocks from disks into memory in two stages. During initialization, \NAME{} identifies and reads into memory code blocks with significant differences. In the runtime stage, subsequent blocks are asynchronously pre-read into memory after action matching.

Our evaluation shows that the latency of \NAME{} is 45.1\% to 75.8\% lower than native Linux based on SD card and up to 67.7\% lower than that of NFS in wired environments. For real-world applications in wireless environments, the performance of \NAME{} is at best only 22.9\% worse than that of the local Linux, which is almost comparable with local Linux and surpasses NFS by up to 60.7\%. The power consumption of \NAME{} is at worst 6.7\% higher than NFS, but it can save energy consumption due to the fact that \NAME{} can load executables with a greater network bandwidth continuously and consistently.

The contributions of our work are summarized as follows:
\begin{itemize}
    \item We design and implement \NAME{}, a lightweight executable delivery system based on Linux at the block layer. This system allows applications to execute efficiently by loading necessary blocks for executables and their dependencies during runtime, eliminating the need for complete pre-downloads. It can avoid downloading complete executables and dependencies to local storage.
    \item We propose an action-based block stream prefetching mechanism based on the history of access patterns. The server can accurately predict and deliver required code blocks to clients by abstracting these patterns into actions.
    \item We develop a normalized-variance-based method to identify and pre-load block sequences with significant differences from disks to memory during the initialization stage. Additionally, \NAME{} can asynchronously prefetch subsequent blocks in the matched action during runtime to avoid blocking.
    \item We demonstrate that the latency of \NAME{} is up to 75.8\% lower than that of native Linux based on SD card and 67.7\% lower than that of NFS in wired environments. Even for real-world applications in wireless environments, \NAME{} outperforms NFS by up to 60.7\% and achieves a comparable performance with local Linux. The power consumption of \NAME{} is at worst 6.7\% higher than NFS, but it can save energy consumption due to a lower processing time.
\end{itemize}

%% file: 2_background.tex
\section{Observations}{\label{background}}
We motivate \NAME{} by analyzing the IO overhead of IoT devices in different ways (§\ref{sec-perf-blockio}).
Then, we scrutinize the Linux read-ahead algorithm and propose our action-based prefetch mechanism (§\ref{sec-prefetch}).

\begin{table}
    \centering
    \caption{Breakdown of block IO}
    \setlength{\tabcolsep}{0.9mm}
    \footnotesize
    \label{tab-breakdown}
    \begin{tabular}{c c c | c c c c c c c}
        ~ & \multicolumn{2}{c}{\textbf{Ext4(SD)}}& 
        ~ & \multicolumn{2}{c}{\textbf{NFS4}} &  \multicolumn{2}{c}{\textbf{Prototype}} \\
        Module & us	& \% & Module & us & \% & us	& \% \\
        \hline
        Mapping & 11.5 & 1.8 & Mapping & 15.1 & 2.4 & 6.4 & 1.2 \\
        Gen Block & 9.1 & 1.5 & Middle & 6.3 & 1 & 7.1 & 1.3 \\
        \hline
        IO Sched & 13.7 & 2.2 & \multirow{3}{*}{\makecell{IO\\(Net/Disk)}} & \multirow{3}{*}{604.5} & \multirow{3}{*}{96.6} & \multirow{3}{*}{507.3} & \multirow{3}{*}{97.5}\\
        Driver & 53.3 & 8.7 & ~ & ~ & ~ & ~ & ~ \\
        Device & 527.6 & 85.8 & ~ & ~ & ~ & ~ & ~ \\
        \hline
        Total & 615.2 & 100 & Total & 625.9 & 100 & 520.8 & 100 \\
    \end{tabular}
\end{table}

\subsection{The performance of block IO}{\label{sec-perf-blockio}}

To analyze the performance of different block IOs for accessing code blocks, 
we used a Raspberry Pi as the client to run Python and measure the average latency of accessing a single page. 
The latency calculation, commencing and concluding at the Mapping Layer beneath the Virtual File System (VFS), encompasses the time from disk IO interception to completion.
The results are summarized in Table \ref{tab-breakdown}, which provides a breakdown of a single block IO request. 
The breakdown mainly includes the Mapping Layer and the IO Layer. 
We evaluated local Ext4, NFS, and a lightweight prototype.

\begin{figure}
    \centering 
    \includegraphics[width=0.4\textwidth]{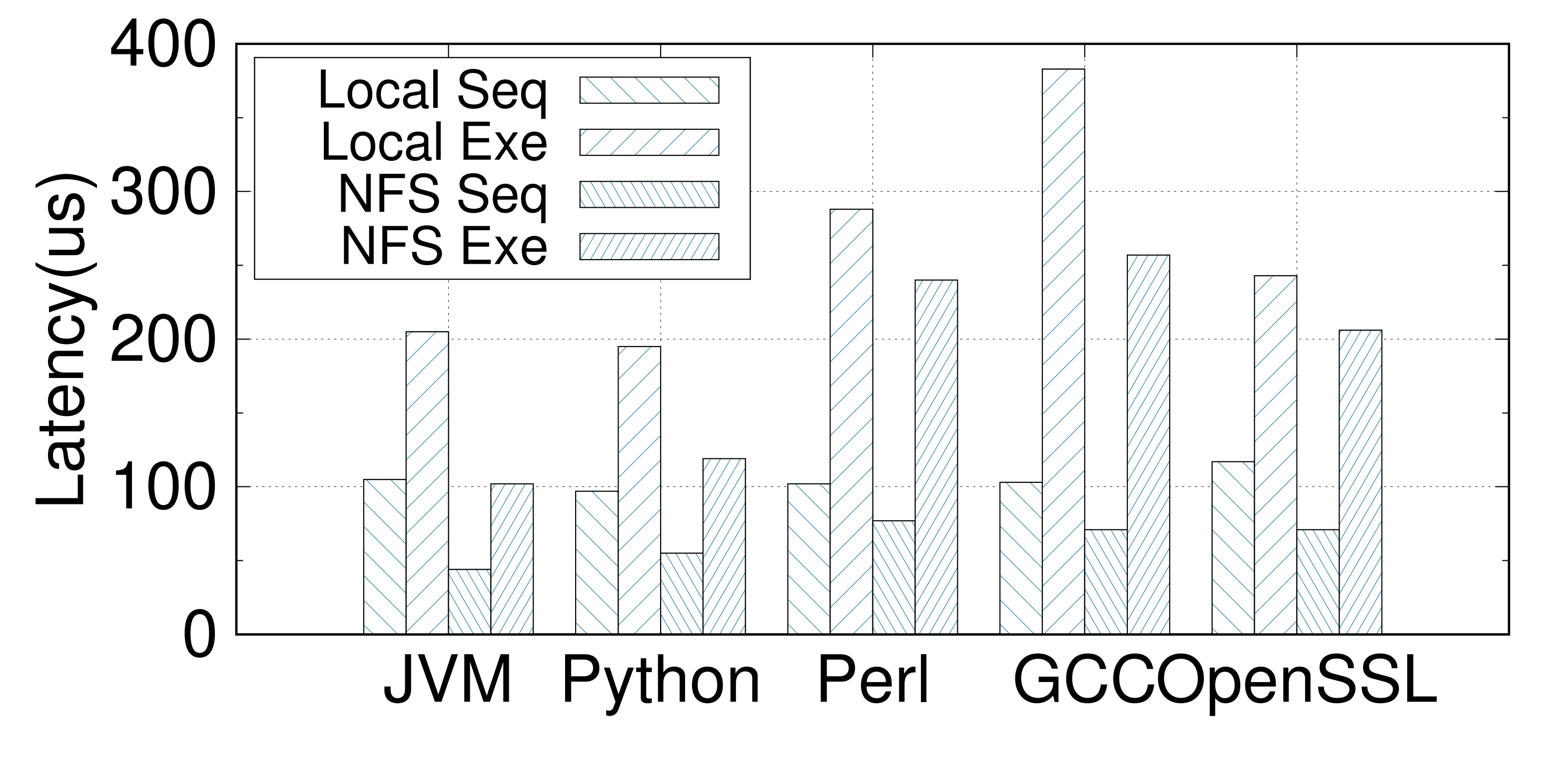}
    \vspace{-0.4cm}
    \caption{Latency for reading a single page using local Linux filesystem and NFS, in both sequential and execution order.} 
    \label{img-seq-exec} 
    \vspace{-0.5cm}
\end{figure}

\textbf{Local block IO}. Linux optimizes local block IO, with the Mapping and Generic Layers accounting for around 20\% of the total IO overhead and the IO scheduling layer and driver layer making up 10\%. The majority of overhead (72\%) lies in storage device response time, hence the importance of accurate prefetching to populate the Page Cache.

\textbf{Complex NFS}. NFS\cite{nfs} enables network file and directory access via multiple threads like \textit{rpc}, \textit{nfsd}, and \textit{mountd}, yet at high costs. Its read-ahead algorithm is efficient for sequential/stride data files but falls short for executables, triggering frequent disk reads. NFS also maintains file and directory status for client-server consistency, incurring more network round-trips. Consequently, NFS's IO Layer overhead stands at 604.5us, or 96.6\% of total overhead, covering client's network IO and server's disk IO overhead.

\textbf{Lightweight prototype}. The prototype system of \NAME{} embodies the application delivery framework, employing lightweight IO optimization (§\ref{sec-opt-iopath}). It uses the same native Linux read-ahead algorithm as NFS but with lower overhead. Focused on network-based file reading over supporting file operations and complex state maintenance, the prototype's IO Layer overhead is close to that of local block IO. It performs 16.8\% better than NFS due to NFS's complexity and higher IoT device burden.


\subsection{Prefetching}{\label{sec-prefetch}}

\begin{table}
    \centering
    \caption{Pre-Read analysis of various applications.}
    \setlength{\tabcolsep}{0.8mm}
    \footnotesize
    \label{tab-pre-read}
    \begingroup
    \setlength{\tabcolsep}{5pt}
    \begin{tabular}{c | c c c c c c c c}
        ~ & T & N & P & N/T & N/P & P/T & IO & B/IO\\
        \hline
        JVM & 2803 & 1651 & 2720 & 59\% & 60\% & 97\% & 117 & 23.2 \\
        Python & 1149 &	519	& 883 &	45\% & 59\%	& 77\% & 38 & 23.2 \\
        Perl & 782 & 408 & 634 & 52\% & 64\% & 81\% & 28 & 22.6 \\
        GCC & 269 & 97 & 269 & 36\% & 36\% & 100\% & 13 & 20.6 \\
        OpenSSL & 131 & 63 & 131 & 48\% & 48\% & 100\% & 7 & 18.7 \\
    \end{tabular}
    \endgroup
    \vspace{-0.5cm}
\end{table}

In computers, prefetching can work at the hardware or software level.
At the hardware level, prefetching typically pre-read to cache, or TLB \cite{jain2013linearizing,michaud2016best,hashemi2018learning,ayers2020classifying,jalili2022managing}. 
This procedure involves analyzing the current and historical access pattern and pre-reading data from the predicted subsequent address to TLB or cache in advance. 

Software prefetching \cite{fengguang2008design, readahead, al2020effectively, cao1994implementation, griffioen1994reducing, kaplan2002adaptive}, works in a similar way to hardware prefetching, aims to mitigate the speed mismatch between memory and block devices. 
The stream-prefetching-based Linux read-ahead algorithm \cite{readahead}  pre-reads the subsequent file blocks with varying lengths from disk to memory ahead.
Leap\cite{al2020effectively}, which learns from hardware prefetchers, captures the offset differences as access patterns and leverages majority trend-based prefetching to resist potential disturbances.
However, they're optimized for sequential and stride access patterns in data files, and thus less efficient for unpredictable access patterns in executables due to complex instructions. 
Figure \ref{img-seq-exec} illustrates this predicament, showing that reading executable files in execution order, compared to sequential reading, incurs longer processing times (1x-2.7x in local Linux, 1.1x-2.6x in NFS). 
This leads to the relative inefficacy of existing solutions based on the Linux native read-ahead algorithm for executables.

To address this, we analyzed various applications' execution, gathering key statistics summarized in Table~\ref{tab-pre-read}. The data revealed that while only 36\%-59\% of code blocks are accessed (Need/Total), an aggressive 77\%-100\% are pre-read (Pre-read/Total), causing unnecessary disk IO.
The analysis further showed that despite the illusion of a 100\% hit rate for GCC and OpenSSL achieved through aggressive read-ahead, the actual performance of GCC is impaired due to its low N/T ratio.
GCC suffered due to its low N/T ratio, while OpenSSL performed better due to its smaller binary size. 
Best performance is seen with Python and JVM due to their large binary size and high Blocks/IO ratio (B/IO), allowing for effective pre-reading. 
This demonstrates a need for a more efficient read-ahead algorithm and a more accurate pre-reading method to minimize waste (P/T).

\begin{figure}
    \centering 
    \includegraphics[width=0.48\textwidth]{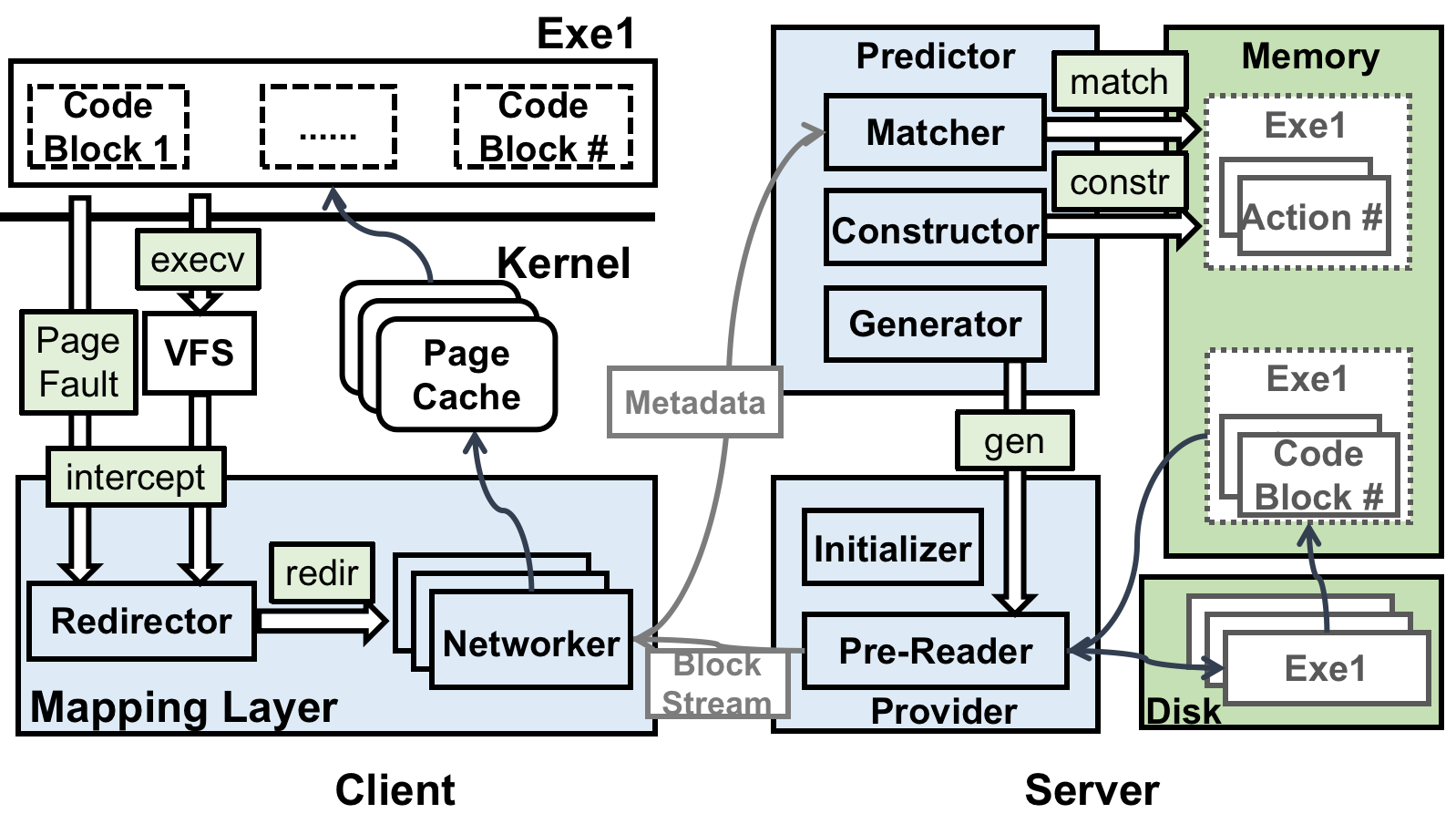}
    \caption{System overview of \NAME{}} 
    \label{img-Architecture} 
    \vspace{-0.6cm}
\end{figure}

These findings indicate a need for a more efficient read-ahead algorithm, especially for handling irregular access patterns.
Considering the moderate sizes of executables, overly complex prediction algorithms could induce high latency. 
Typically, applications request blocks of executables as a continuous code block stream.
During the execution of a task, a block stream is requested within a period, demonstrating continuity. 
Our observations show consistency between requested block streams under different parameters.
In IoT devices, tasks are frequently repeated to complete a function, causing the application to be executed multiple times at the granularity of the task and exhibiting repeatability.
These characteristics call for utilizing historical knowledge for predictions, which offers a beneficial approach to provide great performance(as discussed in §\ref{sec-action}). 

In summary, when delivering executables via the network, network IO and disk IO on the server account for the majority of overheads. 
Thus, optimizing network IO, such as transitioning from synchronous to asynchronous operations, and enhancing disk IO, for instance, maximizing caching of executables in memory, become crucial aspects to focus on. 
It's also important to boost the accuracy of executable prefetching, which can reduce network IO round trips and improve the B/IO parameter to expand each IO's payload. 
This necessitates a trade-off between employing complex yet precise prediction mechanisms and considering the associated cost of prediction.



%% file: 3_design.tex
\section{System Design}
\label{design}
This section presents the core design of \NAME{}. \textbf{Low IO overhead} and \textbf{Precise prefetching} are two main goals of \NAME{}.
Given that the size of executables is typically not overly large (usually in the order of tens of M bytes), there is no need for a particularly complex algorithm to handle them. 
Therefore, adhering to the principle of Occam's Razor, we have designed a concise predictor based on history, ensuring good performance.

Comprising both the client and server, \NAME{} utilizes three key components on the client side (Redirector, Networker, and Allocator) and two on the server side (Predictor and Provider). 
The client components redirect local disk IO to network IO and handle received code block streams, while the server components predict, pre-read, and send possible subsequent block streams back to the client. 
The architecture of \NAME{} is depicted in Figure \ref{img-Architecture}.

\begin{figure}
    \centering 
    \includegraphics[width=0.48\textwidth]{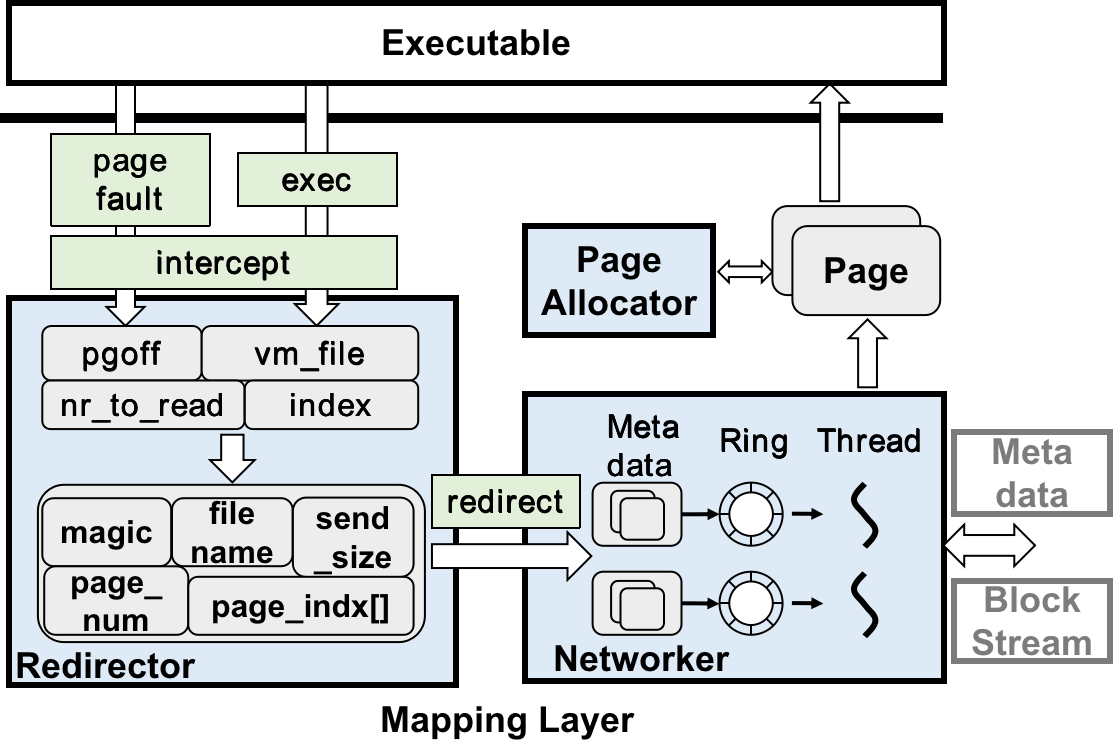}
    \caption{Optimised IO path} 
    \label{img-iopath} 
    \vspace{-0.3in}
\end{figure}

\subsection{Optimised IO Path}{\label{sec-opt-iopath}}
When a client initiates an application and the required code is not loaded into memory yet, the \textit{exec} system call activates the kernel's ELF loader, triggering an IO request to the local filesystem for the executable's initial code block, which includes the main function. 
As this block and subsequent ones execute, they generate a series of page faults based on the code path, leading to a corresponding series of IO requests.

\textbf{Redirector}. 
When an IO request reaches the Mapping Layer, the Redirector intercepts it. 
This disk IO, containing the file (executable) name and page (block) index, aids disk drivers in locating the code block. 
However, as the executables are stored on the server, the normal disk IO path is redirected to network IO by the Redirector.
To accomplish this, the Redirector maintains a configurable set of executable names, stored in a hash table.
Simultaneously, the server holds the corresponding executables for the names kept within this set. 
Whenever the Redirector intercepts a disk IO at the entry of the mapping layer (e.g., within the \textit{do\_fault} kernel function), it initiates a search process to match the disk IO's file name with the names stored in the executable name set.
If the match fails, it seamlessly proceeds to access the local storage using EXT4. 
Otherwise, it performs a redirection operation.
The Redirector encapsulates various pieces of information, including but not limited to the file name and page index, as metadata in a specific format that \NAME{} on the server side can recognize. 
After storing the metadata in a ring shared with the Networker, the Redirector initiates the redirection of disk IO by changing the IO path.

\textbf{Networker}. 
Once the Redirector redirects local disk IO, the Networker dequeues metadata from the ring, transmits it and receives the corresponding code block stream from the server using TCP. 
Upon processing the request, the server-side \NAME{} returns the code blocks and their quantity. 
The Networker subsequently allocates an equivalent number of free pages from the Allocator for storing the blocks.

However, dealing with a large volume of instantaneous IO requests can often lead to frequent connection establishment and release, resulting in a blocked IO path. 
To overcome this issue, we remove network IO from the main kernel path and implement it as an independent agent. 
This approach affords two significant benefits.
First, by functioning as an independent entity, the Networker can maintain persistent connections, efficiently handling sudden surges of network requests. 
This method provides non-blocking network connection services, thereby eliminating the overhead associated with establishing and releasing a connection for each IO request.
Secondly, since each CPU core generates disk IO, Networker creates a private thread for each core, enhancing the concurrency of block IO.

\textbf{Allocator}.
To optimize memory space and alleviate overhead during page allocation, we introduce a separate component called the Allocator within the Networker. This design ensures that allocation tasks do not obstruct the Networker's operations. The Allocator, functioning as an independent thread, asynchronously manages a page pool organized as a linked list with an empirically set size of 256. Employing a producer-consumer model, the Networker acts as the consumer, obtaining free pages from the pool and informing the Allocator about the number of pages required. In turn, the Allocator, as the producer, requests the corresponding pages from the kernel and stores them in the page pool. As the Networker fills these free pages with code blocks received from the server, the Allocator adds them to the Page Cache, allowing continuous execution of the executable without local storage installation. In cases where sufficient client memory is available, the pages of the executable in the Page Cache are retained and managed by the kernel using a Least Recently Used (LRU) mechanism.

\subsection{Action-based Predictor}{\label{sec-action}}

\begin{figure}
    \centering 
    \includegraphics[width=0.32\textwidth]{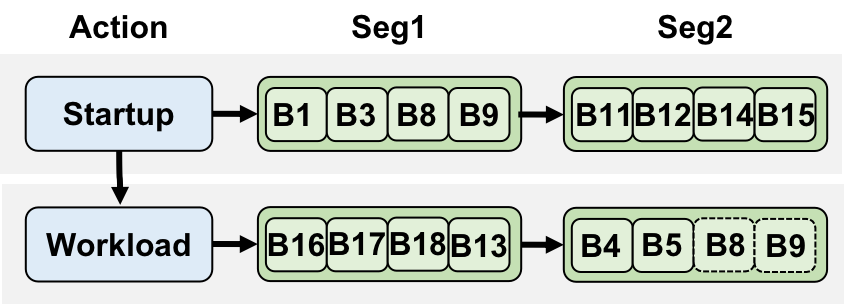}
    \caption{A case of an action, segment and block.} 
    \label{img-act-seg-blk} 
    \vspace{-0.2in}
\end{figure}

Observing IoT device executable access patterns (§\ref{sec-prefetch}), we abstract task code block streams into actions. Each action encapsulates a sequence of requested code blocks, representing the expected code path and aiding in task predictions. Interactive applications typically exhibit three action types: Startup, Exit, and Workload(s), while batch applications have a single action encompassing these three.
Given consistent application executions and relatively small executable sizes, tracking and matching code block streams with low overhead at a 4K size block granularity is feasible. Utilizing a prediction approach based on execution history provides an effective trade-off between complex, precise predictions, and associated overhead.


As shown in Figure \ref{img-Architecture}, the Prefetcher is a two-part system encompassing the Predictor and Provider. 
The Predictor anticipates the code block stream needed by the executable, while the Provider fetches these blocks from the disk or memory and sends them to the client. 
Informed by our analysis of page access patterns, we've developed an action-based Predictor. 
This component constructs the action, matches it to the block stream, and then generates the corresponding sequence of block streams for the application task.

\begin{figure*}[!htb]
	\centering
	\begin{subfigure}[t]{0.32\linewidth}
		\begin{minipage}[t]{1\linewidth}
            \includegraphics[width=1\linewidth]{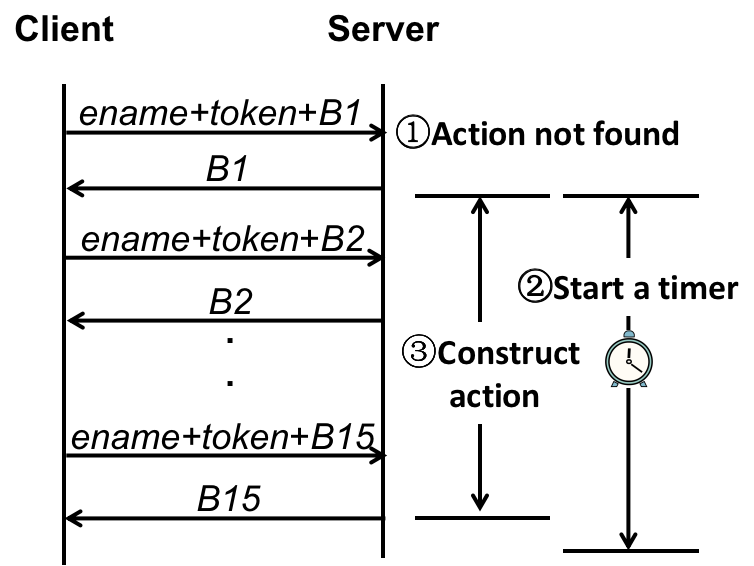}
		\end{minipage}
		\caption{Action Construction}
        \label{action-constr}
	\end{subfigure}
        \begin{subfigure}[t]{0.32\linewidth}
		\begin{minipage}[t]{1\linewidth}
			\includegraphics[width=1\linewidth]{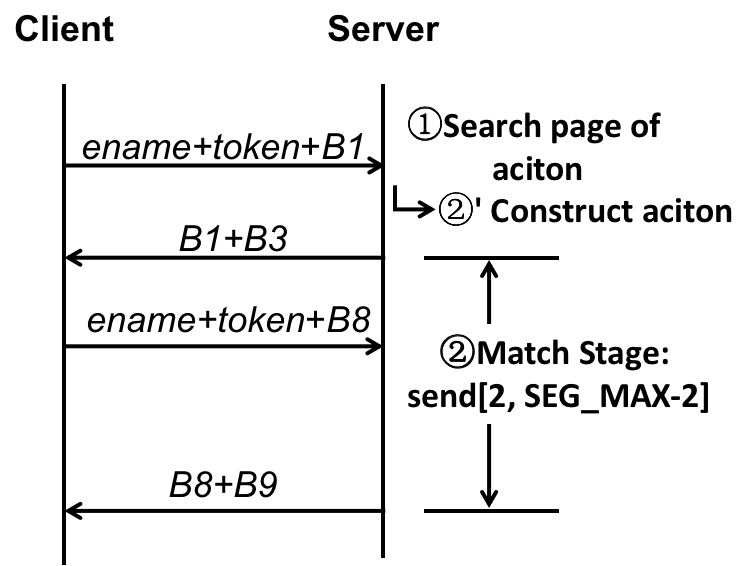}
		\end{minipage}
		\caption{Action Match}
        \label{action-match}
	\end{subfigure}
	\begin{subfigure}[t]{0.32\linewidth}
		\begin{minipage}[t]{1\linewidth}
		  \includegraphics[width=1\linewidth]{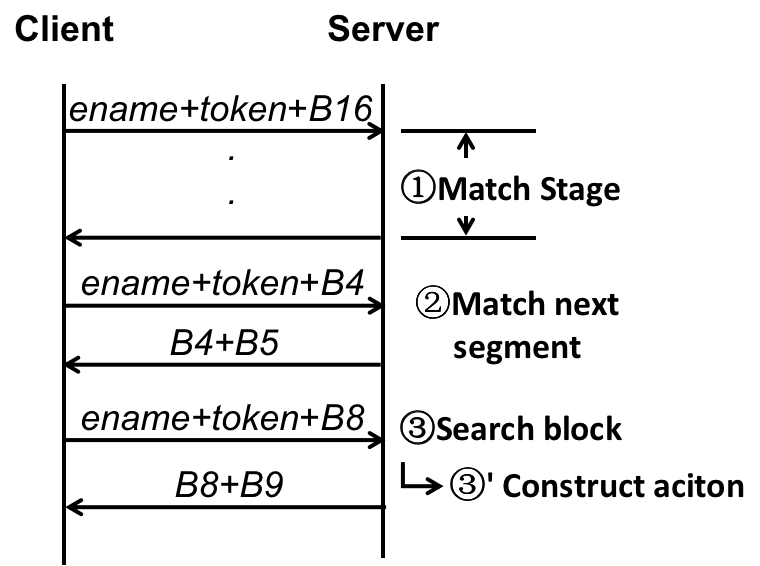}
		\end{minipage}
		\caption{Action Generation}
        \label{action-gen}
	\end{subfigure}
	\caption{Workflow of the action-based prefetcher.}
	\label{action}
        \vspace{-0.28in}
\end{figure*}

As shown in Figure \ref{img-act-seg-blk}, the server-side \NAME{} maintains multiple actions for each executable, including starup, exit and workload(s).
Each action is organized by one or more segments using a linked list, with each segment containing a sequence of code block indexes (with an experience value of 32).
Segments are used to divide the code blocks stream so that the Predictor and Provider can control the granularity of prefetching and pushing code blocks to the client conveniently.
When the Predictor makes predictions based on actions, the prediction granularity is one segment, indicating that Provider provides the client with one segment at a time.
The workflow of the Predictor is illustrated in Figure \ref{action}. 
\NAME{} maintains tokens to an action in a many-to-one manner, with each token representing an executing action.
Server-side \NAME{} maintains the action state based on the combination of token and executable name, which includes the stage it is in, the segment it requires next time, the segments it prefetches, and so on.
Each token is a UUID generated by Networker for every successful match of a new executable name with an action. The token expires once the corresponding action is completed.
In this section, we will use actions in Figure \ref{img-act-seg-blk} as examples.

\textbf{Action Construction}. 
When the application executes, it retrieves necessary code blocks from the server where the entire executable is stored. During the initial serving of the executable, the server records the block index of the requested block stream to construct an action. Throughout the Action Construction phase, the client requests the block stream piece by piece, with the Predictor returning one page to the client per network round-trip.


As shown in Figure \ref{action-constr}, the client initially sends the token, executable name, and block index (B1) to the server. The Predictor begins the matching phase to locate the corresponding action for the executable name. If no action is found, it signifies that the server-side \NAME{} has yet to construct an action for the current executable, triggering the construction stage. Given our observation that most actions expire within 2 seconds, \NAME{} sets a 3-second timer for action construction. Within this time frame, the Predictor consistently receives block stream requests, storing the block index to the segment.
Upon receiving the sequence (B1,B3,B8,B9), which reaches the segment size \textit{SEG\_MAX}, the predictor allocates a new segment to accommodate the subsequent sequence (B11,B12,B14,B15). Post\-construction, all actions with their code block sequences are saved to persistent storage, ready to be loaded at \NAME{} startup.

\textbf{Action Match}. 
When the Predictor receives an IO request from a client, it matches the token with its saved tokens to find an action. If no match is found, indicating the block isn't part of an existing action, the Predictor initiates a construction stage for this block and subsequent ones. Conversely, if a token match occurs, it signals an ongoing action, prompting the Predictor to retrieve the action state for continued processing. If the Predictor receives metadata with a new token, it identifies the action based on the executable name.

During the matching stage, the Predictor conducts multiple matches for the initial segment to determine if the requested block stream matches the action pattern. 
As shown in Figure~\ref{action-match}, when the Predictor receives the first request B1, it matches the segment \#1 that contains B1 and sends (B1, B3) to the client. 
Throughout this stage, the Predictor maintains a matching pointer, set to target block B8 in the next iteration. Upon receipt of the second request B8 (the request for B3 is not sent to the server as it's already cached on the client), the Predictor correlates the requested block with the match pointer and delivers \textit{SEG\_MAX}-2 blocks to the client, including (B8, B9). To minimize the number of matches and network overhead, multiple matching operations are reserved for the first segment only.

When \textit{SEG\_MAX} exceeds 32, the time of multiple matches can be increased to three, enhancing the precision of action matching. 
Upon the successful matching of the indices at the 2nd and (\textit{SEG\_MAX}-2)th positions within the first segment, the Predictor presumes a successful match between the requested block stream of the executable and the action pattern, thus initiating the generation stage.
Our observations indicate that most action sequences are stable, typically making a match within the first segment sufficient. 
However, when the same application runs concurrently, it may introduce interference that could lead to multiple match failures when relying on a fixed position match pointer. 
In such cases, the Predictor scans through all segments of all actions to locate the probable action and its segment to which the requested block belongs, before sending that segment to the client. 
If no existing action is associated with this block, a new construction stage is launched.

\textbf{Action Generation}. When the match stage matches a certain existing action, it enters the generation stage, as illustrated in Figure \ref{action-gen}.
To avoid traversing each segment, the Predictor maintains the latest segment corresponding to the action by the token.
After the multiple matching in the first segment, such as the (B16,B17,B18,B13) for the \textit{workload} action, the Predictor updates the match pointer to the next segment B4.
Based on the stability feature of the page access pattern, subsequent segments after the match stage only require a single match to speed up the action search. 
In this case, \NAME{} will only match the first block index (B4) of the latest segment to reduce the network round trip.

If the next received request is not B4(e.g., B11) in the context of the \textit{workload} action, it indicates that the generation stage has failed to find a match.
This may occur when the executable has jumped to a new execution path based on the previous block stream, typically caused by a thread switch.
In this case, the Predictor needs to start a new match stage for B11.
If the match is successful, a new action will be used to provide block request services.
Otherwise, a new action will be constructed for the new block stream.

\subsection{Provider}{\label{sec-provider}}

\begin{figure}
    \centering 
    \includegraphics[width=0.49\textwidth]{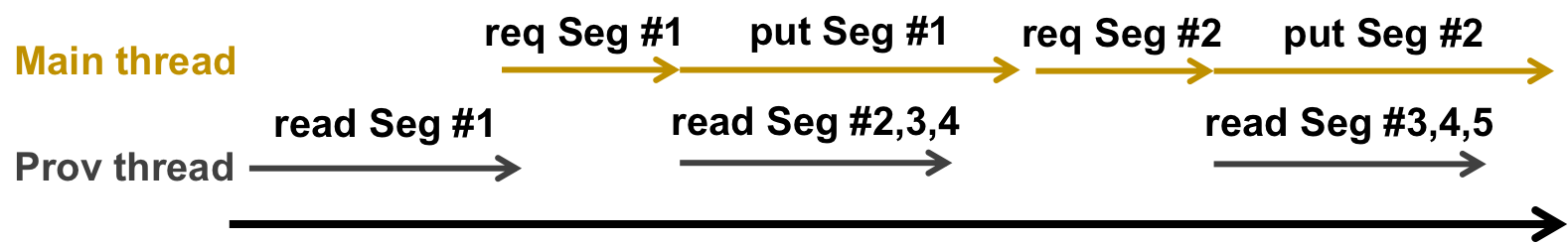}
    \caption{An example of asynchronous prefetch.} 
    \label{img-async-prefetch} 
   \vspace{-0.28in}
\end{figure}

When a client initiates a block request, the server-side \NAME{} reads the corresponding block from the disk and sends it back to the client. Unfortunately, disk IO is a bottleneck in the entire IO path, limiting performance.
To alleviate this issue, we must preload the executables into memory ahead of time to reduce disk IO overhead. 
However, due to memory size constraints on the server, it's unfeasible to load all executables into memory. 
Therefore, we must selectively load only the necessary parts of the executables into memory. Instead of relying on multi-level cache systems, which can be complex and inefficient, we seek a more efficient approach to address this problem.


\textbf{Pre-read during Initialization.} The Provider calculates the variance for all segments of the executable to represent the difference in the code block sequence of various segments.
Since different applications have different sizes, their code block index values also vary.
To standardize the measurement of variance, the Provider first calculates the normalized value of each block index in each segment before calculating the variance. 
The normalized value is calculated using the formula $N_i=(B_i-B_{min})/(B_{max}-B_{min})$, where $B_i$ represents the index of the \textit{i-th} block.
Next, the Provider calculates the variance of segments based on the normalized values. 
It calculates the average normalized value of the segment as $avg_{seg}$, and uses the formula $sum=(avg_{seg}-N_i)*(avg_{seg}-N_i)$ to calculate the variance of the segment.
The Provider then compares the normalized variance value of each segment with a threshold. 
Segments that exceed the threshold will be read into memory during the initialization stage.

\textbf{Pre-read during Runtime.} After the Provider pre-reads the segments based on the normalized variance in the initialization phase, it asynchronously pre-reads the segments that the client will access subsequently during the runtime phase.
The server-side \NAME{} creates a Prov thread during the initialization phase.
Although it pre-reads segments with high normalized variance during initialization, the Predictor cannot predict when applications will run. It still faces the first segment dilemma, which brings significant overhead for the first match.
To enhance the performance of fetching the first segment of each action, after the Prov thread is created, \NAME{} traverses the first segment of each executable's action and reads each page saved by the first segment. 
 As shown in Figure \ref{img-async-prefetch}, when the server receives the request of segment \#1, the main thread reads pages of segment \#1 directly from the server's page cache. 
Meanwhile, the Prov thread reads segment (\#2,\#3,\#4) asynchronously to prepare for the subsequent application's access to segment \#2, and so on.

%% file: 4_implementation.tex
\section{Implementation}{\label{implementation}}
The \NAME{} we implemented spans both the client and server sides, with the client code running on the Raspberry Pi4B and the server code running on DELL R730.
The client-side \NAME{} runs in kernel mode, where the Redirector contains 188 lines of C code (LoC), and the Networker consists of 477 LoC.
The server-side \NAME{} runs in user mode and is implemented from scratch in 1517 LoC. This includes 650 LoC for the Predictor and 365 LoC for the Provider.

\textbf{Client-Side \NAME{}}. The Redirector intercepts the disk IO request from VFS and obtains metadata information. The metadata is then enqueued to a kfifo connected to the Networker, and a semaphore will notify the Networker to process. Once the Networker has received the block stream, \NAME{} will return to the kernel path.
The Networker is a Linux kernel module inserted into the kernel using \textit{insmod}. 
It creates threads equal to the number of CPU cores and uses the kernel socket to establish a persistent TCP connection with the server \NAME{}.
After the Networker dequeues metadata from kfifo, it sends requests to the server and waits for the returned blocks. 
The Networker applies for a specified number of free pages from the Allocator and receives the block stream returned by the server through the socket.

\textbf{Server-Side \NAME{}}. The server-side \NAME{} waits and parses the client's disk request through the socket and passes the request to the Predictor.
The Predictor saves the mapping between executables and actions via a hash table, enabling it to quickly retrieve the correct action even if there are multiple executables.
Meanwhile, to avoid the overhead of traversing the current action, the Predictor maintains a data structure storing the segment that the current action will use next time.
The Provider calls \textit{mmap} for all executables during initialization to facilitate access to file content via memory addresses.
The Provider calculates the memory address \textit{(page\_index*PAGESIZE)} of the page during the initialization and runtime stage and then reads corresponding blocks from disks to the page cache.
In this way, the Predictor can read the blocks from the page cache asynchronously.

%% file: 5_evaluation.tex
\section{Evaluation}{\label{evaluation}}


We initially employ microbenchmarks to comprehensively evaluate the network IO and action-based prefetching mechanism of \NAME{}. Subsequently, we evaluate five commonly used real-world application environments to show its performance. Then, we evaluate the performance while running applications in a wireless environment. Lastly, we evaluate the power consumption of it under diverse environments.

\textbf{Testbed.} 
Our evaluations are conducted on a Raspberry Pi 4B and a Dell R730 server. The Raspberry Pi 4B comes with a 1.5GHz Quad-core Cortex-A72 (ARM v8) 64-bit CPU, 4GB LPDDR4 RAM (Rev1.2), and a 16GB Micro SD card. It supports 1 Gigabit Ethernet and 2.4 GHz IEEE 802.11ac wireless connectivity. The R730 server, on the other hand, is equipped with 32 CPU cores, 64GB LPDDR4 memory, and 1 Gigabit Ethernet. In our setup, the Raspberry Pi 4B acts as the client and the R730 server performs the server functions. For a wired configuration, the two devices are interconnected via a 1Gbps Ethernet switch, while for a wireless setup, they are linked via a 1Gbps NETGEAR wireless router.

\textbf{Baseline.} We  run the same binary application on \NAME{}, native Linux, and NFS to make comparisons.
Considering the significant impact of page cache, which can circumvent disk IO upon hits and might affect the subsequent tests, we make sure to clear both client and server cache prior to each test run, unless stated otherwise.

\subsection{Microbenchmark}

We evaluate \NAME{}'s network IO and prefetching performance in microbenchmarks.

\textit{\textbf{Optimized IO Path.}} 
We evaluate the performance of the \NAME{} Prototype using both native network IO and Optimized IO. 
Since our focus is performance enhancement of the Optimized IO path, we load all executable blocks into the server's memory to eliminate interference from disk IO and close action-based prefetching on the server side. 
On the client side, the Linux Readahead algorithm is implemented.
We exclude NFS from the comparison, as its prefetch mechanism makes it an unfair benchmark.
With the Optimized IO path, the client maintains a long connection with the server and allocates free pages asynchronously to store executable blocks from the server. 
In contrast, native network IO establishes short network connections for each request and allocate pages synchronously. 

\begin{figure*}[htbp]
    \centering
    \begin{minipage}[t]{0.73\linewidth}
        \centering
	\begin{subfigure}[t]{0.32\linewidth}
		\begin{minipage}[t]{1\linewidth}
            \includegraphics[width=1\linewidth]{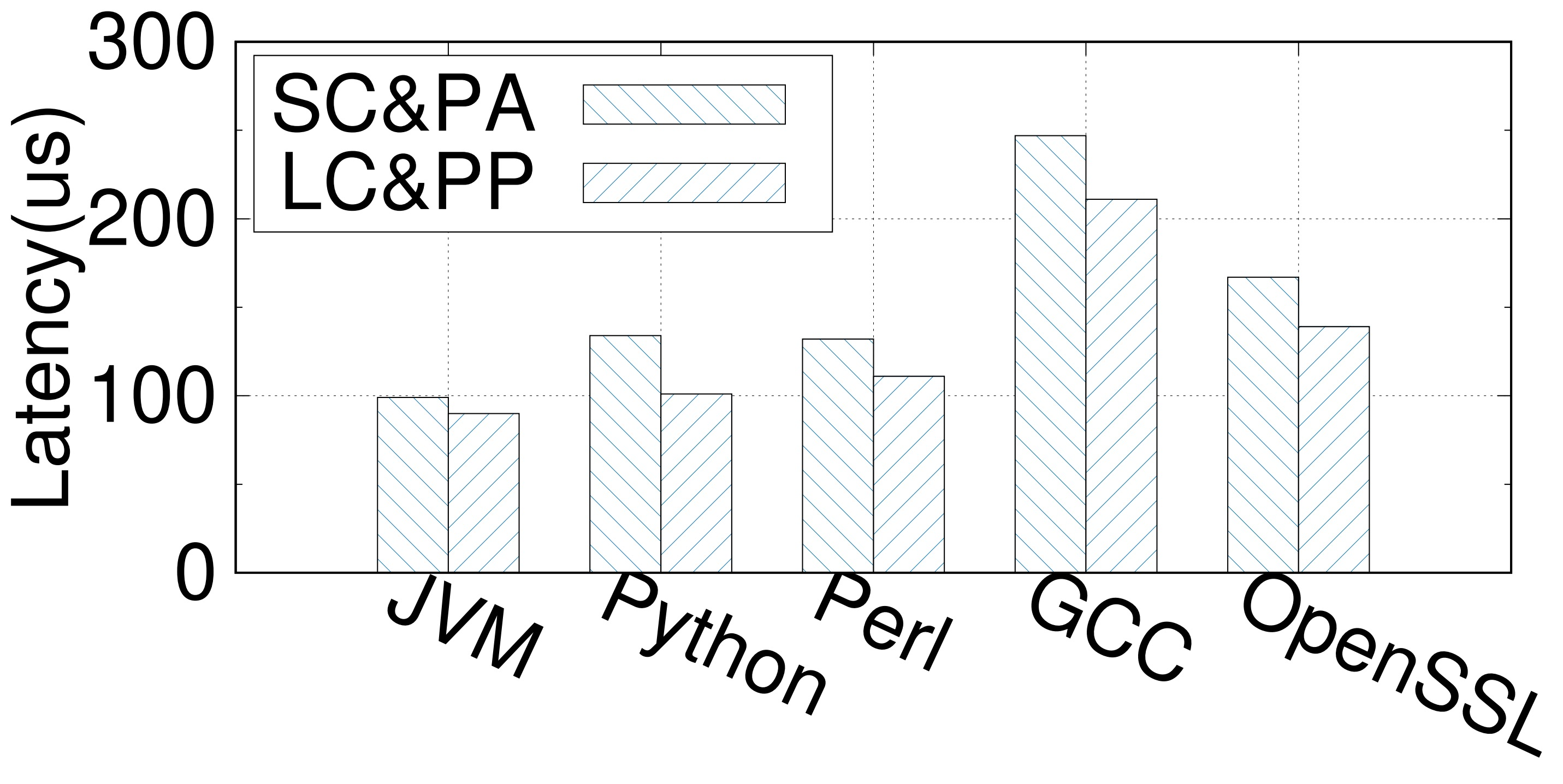}
		\end{minipage}
		\caption{Average latency of different IO path fetching single page.}
        \label{img-optimized-io}
	\end{subfigure}
        \begin{subfigure}[t]{0.32\linewidth}
		\begin{minipage}[t]{1\linewidth}
			\includegraphics[width=1\linewidth]{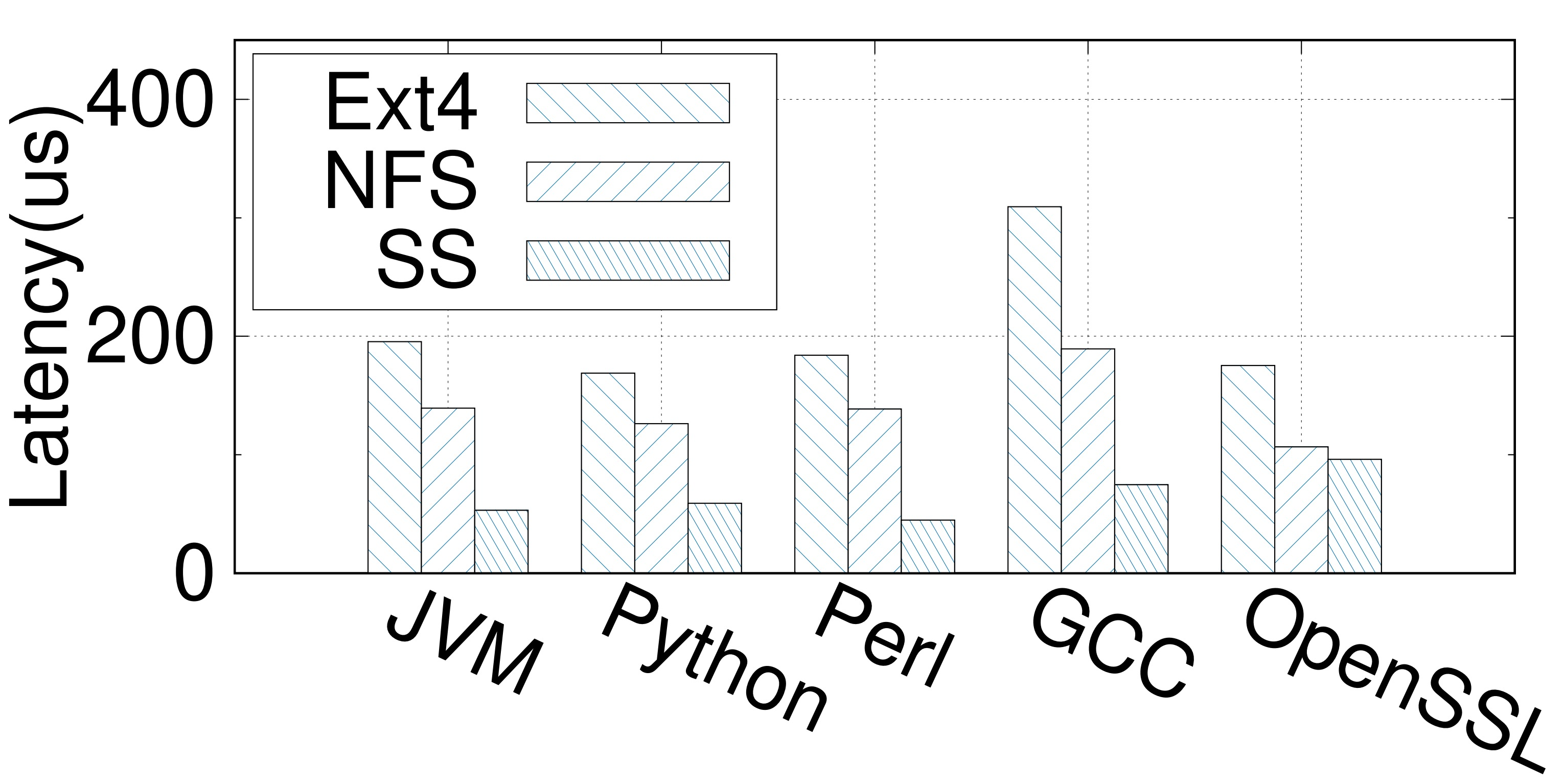}
		\end{minipage}
		\caption{Average latency of different systems fetching single page.}
        \label{img-microbenchmark}
	\end{subfigure}
        \begin{subfigure}[t]{0.32\linewidth}
		\begin{minipage}[t]{1\linewidth}
		  \includegraphics[width=1\linewidth]{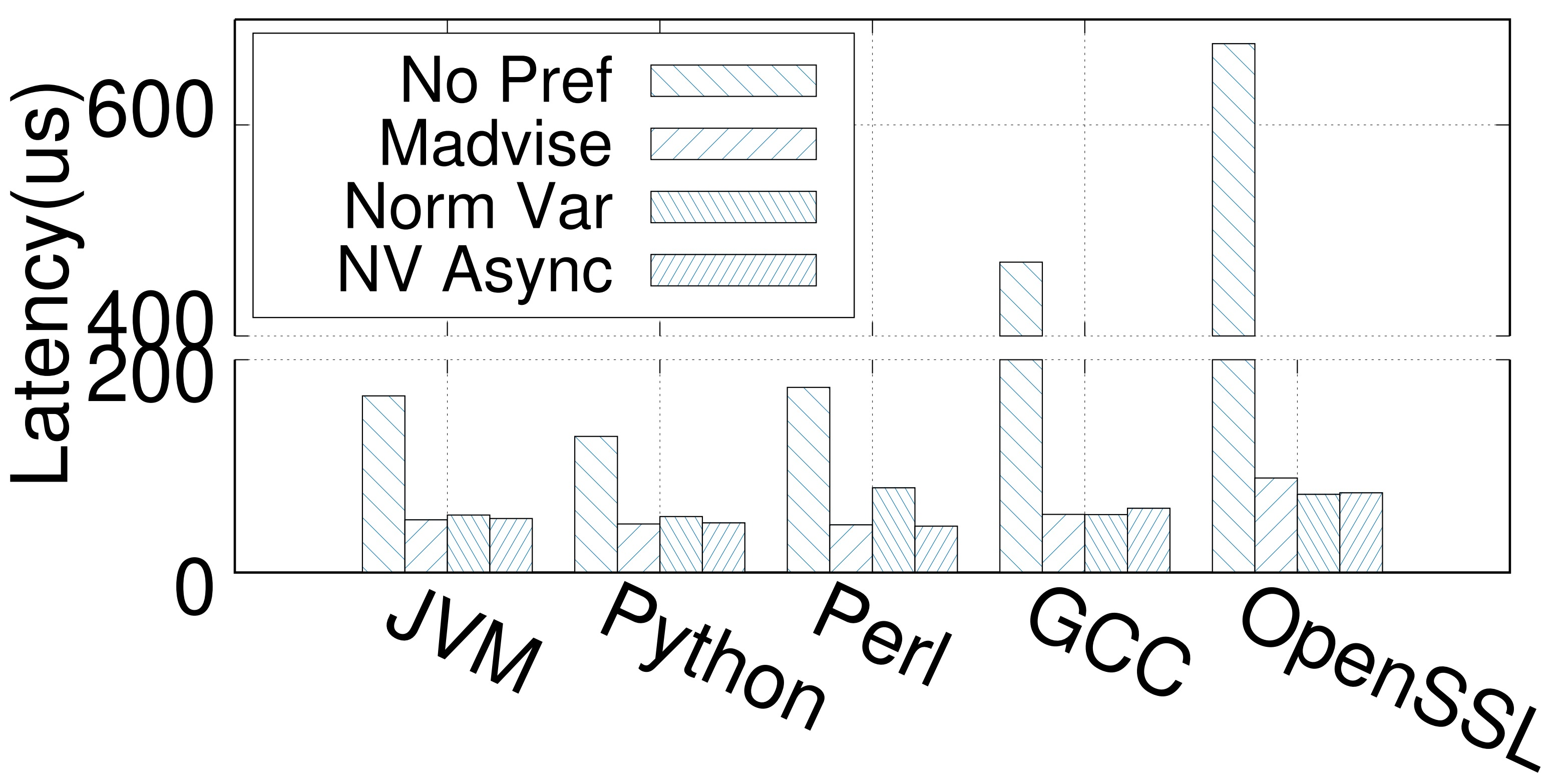}
		\end{minipage}
		\caption{Average latency of different buffering mechanisms.}
        \label{img-provider}
	\end{subfigure}
	\caption{Results of Microbenchmark}
	\label{micro}
    \end{minipage}
    \begin{minipage}[t]{0.23\linewidth}
        \centering
         \includegraphics[width=1\linewidth]{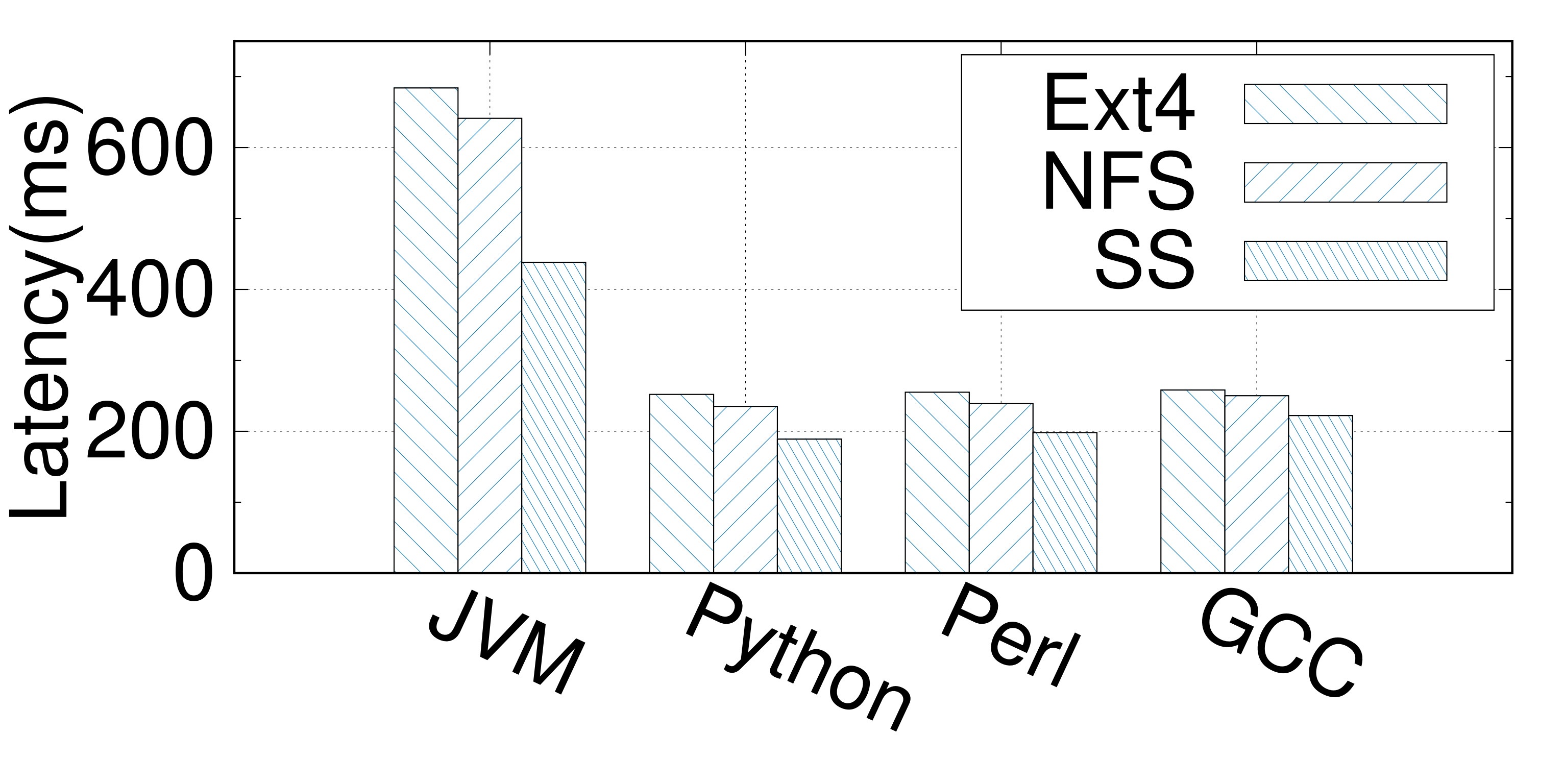}
        \caption{Execution time of applications in macrobenchmarks. } 
        \label{img-macrobenchmark} 
    \end{minipage}
    \vspace{-0.3in}
\end{figure*}

Figure \ref{img-optimized-io} shows that the Optimized IO path, using long connections (LC) and a page pool (PP), enhances performance by 9\%-25\% compared to the naive short connections (SC) and page allocation (PA) approach.
Our observations indicate that the average latency decreases as the number of network IO increases, as more network IO will spread the average cost for each IO.
Though the largest size of JVM executable leads to the most network IO(see table~\ref{tab-pre-read}), it does not benefit much from optimized IO and gains the least performance enhancement.
This is because even under the native network IO path, the code and data for repeated connection establishment and release become hot, allowing the JVM to perform similarly from short connections with long connections. 
In contrast, Python demonstrates the greatest performance improvement as it completes block requests before the code and data of native network IO become hot. 

\textit{\textbf{Prefetching performance.}} To evaluate the overall performance of \NAME{}, we run five commonly used applications and compare their performance with that of native Linux filesystem (Ext4) and NFS. 
 When testing \NAME{} (SS) and NFS, we store the executables on the server, and the client obtains them through the network. 
Figure \ref{img-microbenchmark} shows a comparison of the average latency of different systems accessing blocks. 

The latency of \NAME{} is 45.1\% to 75.8\% lower than that of native Linux based on SD card and up to 67.7\% lower than that of NFS.
According to the results, we find that NFS outperforms Linux based on SD cards because the SD card's speed is even slower than the network. 
\NAME{} achieves better performance than NFS due to its action-based precise prefetching and asynchronous pre-reading, which reduces the number of network round trips and hides disk IO overhead. 
It's worth noting that \NAME{} provides the least performance improvement for openssl, which has a relatively small binary that only requires a few network round trips even with the native read-ahead algorithm of Linux, resulting in less benefit from \NAME{}' accurate pre-reading.


\textbf{\textit{Buffering packets on the server.}} We test the performance of different caching strategies, which play a crucial role in bridging the server-side disk and memory. 
After \NAME{} acquires a block, the code execution typically only consumes a few microseconds. Thus, if the required block is saved on disk, it will block the subsequent code execution for milliseconds.
Figure \ref{img-provider} presents a comparison of \NAME{} performance under different caching strategies, including not using prefetch (No Pref), using madvise (Madvise) to read the whole executable into memory, using the normalized variance (Norm Var) to read segments with higher variances into memory synchronously, and using the normalized variance and asynchronous prefetching (NV Async) strategy to asynchronously pre-read the segment after the current request block.

Results show that server performance with only Linux Read-ahead algorithms (\textit{No Pref}) worsens as executable block count reduces. 
When block count is high, aggressive Read-ahead can preload many blocks into client's memory, it enhances page cache hit probability. 
Yet, for fewer block binaries, the application concludes after a high overhead phase due to more cache misses, resulting in up to 8 times more overhead than NV Async for OpenSSL. 
\textit{Madvise} performs best as it lets \NAME{} hint the kernel about subsequent block access for preload. 
While the normalized variance (\textit{Norm Var}) can preload most large variance segments, it still suffers from imprecise preloading as seen with Python and Perl. 
However, \textit{NV Async}, with asynchronous preloading, effectively reads small variance segments simultaneously, avoiding blocking and achieving similar performance with Madvise. \textit{NV Async}'s latency is 63.5\% to 88.9\% lower than No Pref.

\subsection{Macrobenchmark}{\label{sec-macroben}}
We assess \NAME{}'s performance enhancement capabilities through real-world applications and operations, like running Python/Java applications to remotely load executables like python3/libjvm.so, or compiling a program with gcc.
These applications perform operations like file handling, log printing, and compiling. 
For the sake of testing convenience, the Redirector only matches the main executables 
with the hash table set in the mapping layer, instead of dealing with numerous link files like python byte code.
This setting gives NFS an edge in macrobenchmark evaluation as server-stored files can be accessed faster than from SD cards.

Figure \ref{img-macrobenchmark} shows that \NAME{} performs best, improving performance by 14\%-36\% compared to local Linux Ext4 filesystem. 
Performance improvement is somewhat less in macrobenchmarks due to factors like application and kernel code execution, and context switching. 
Microbenchmarks indicate \NAME{} reduces Python's per page latency by about 110 ms. 
As Python's benchmark needs 519 pages for execution (Table \ref{tab-pre-read}), we estimate \NAME{} offers a performance improvement around 57.1ms (i.e., \textit{110*519/1000}) for Python, consistent with the 63ms test result. 
We omitted OpenSSL from this evaluation due to its benchmark's significant execution time fluctuation caused by the generation of random bytes. 
In conclusion, all applications, especially those with larger executables like JVM, can benefit from \NAME{}.

\begin{figure*}[!htb]
	\centering
	\begin{subfigure}[t]{0.19\linewidth}
		\begin{minipage}[t]{1\linewidth}
			\includegraphics[width=1\linewidth]{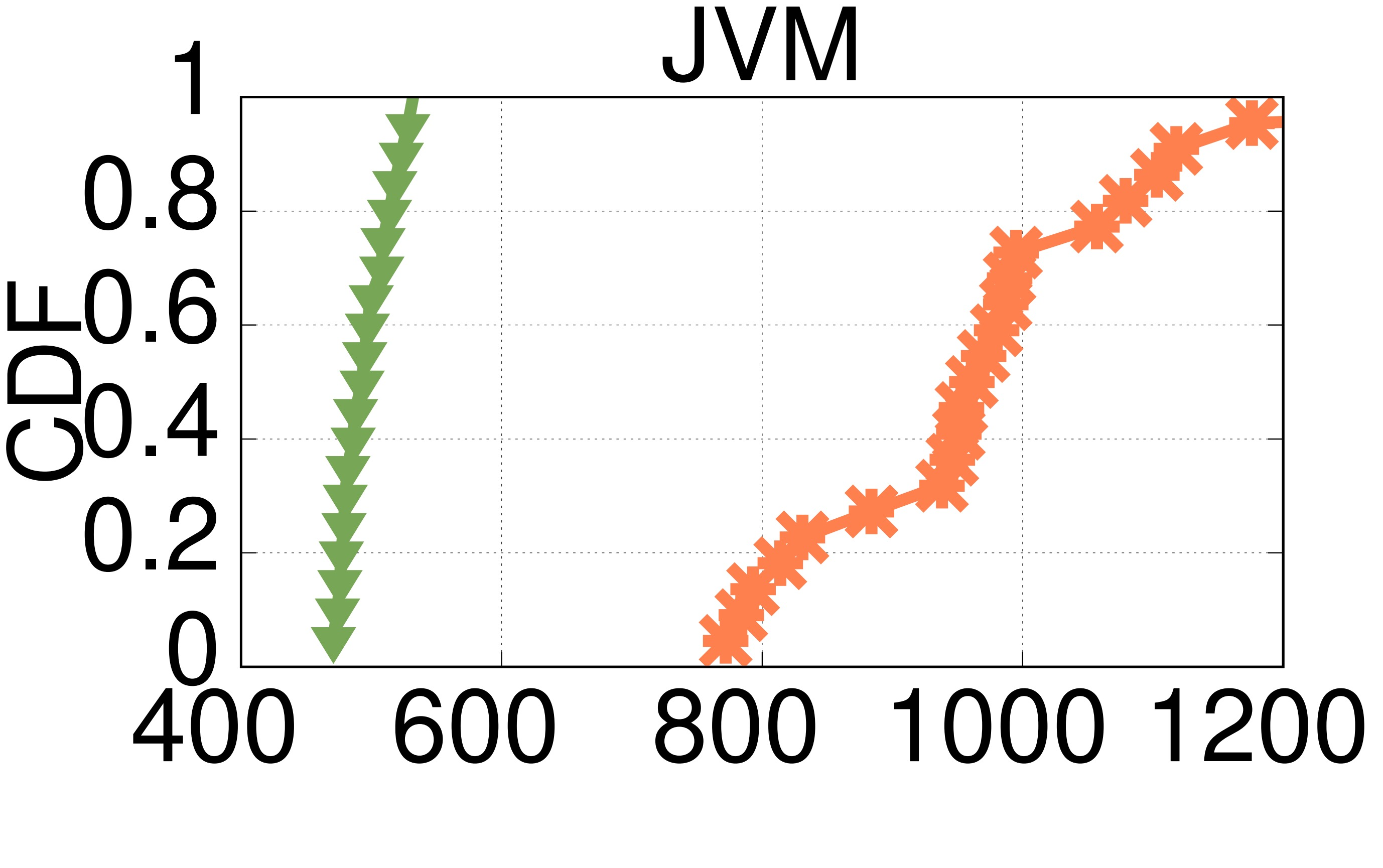}
		\end{minipage}
        \label{img-wifi-micro-jvm}
        \end{subfigure}
        \begin{subfigure}[t]{0.19\linewidth}
		\begin{minipage}[t]{1\linewidth}
			\includegraphics[width=1\linewidth]{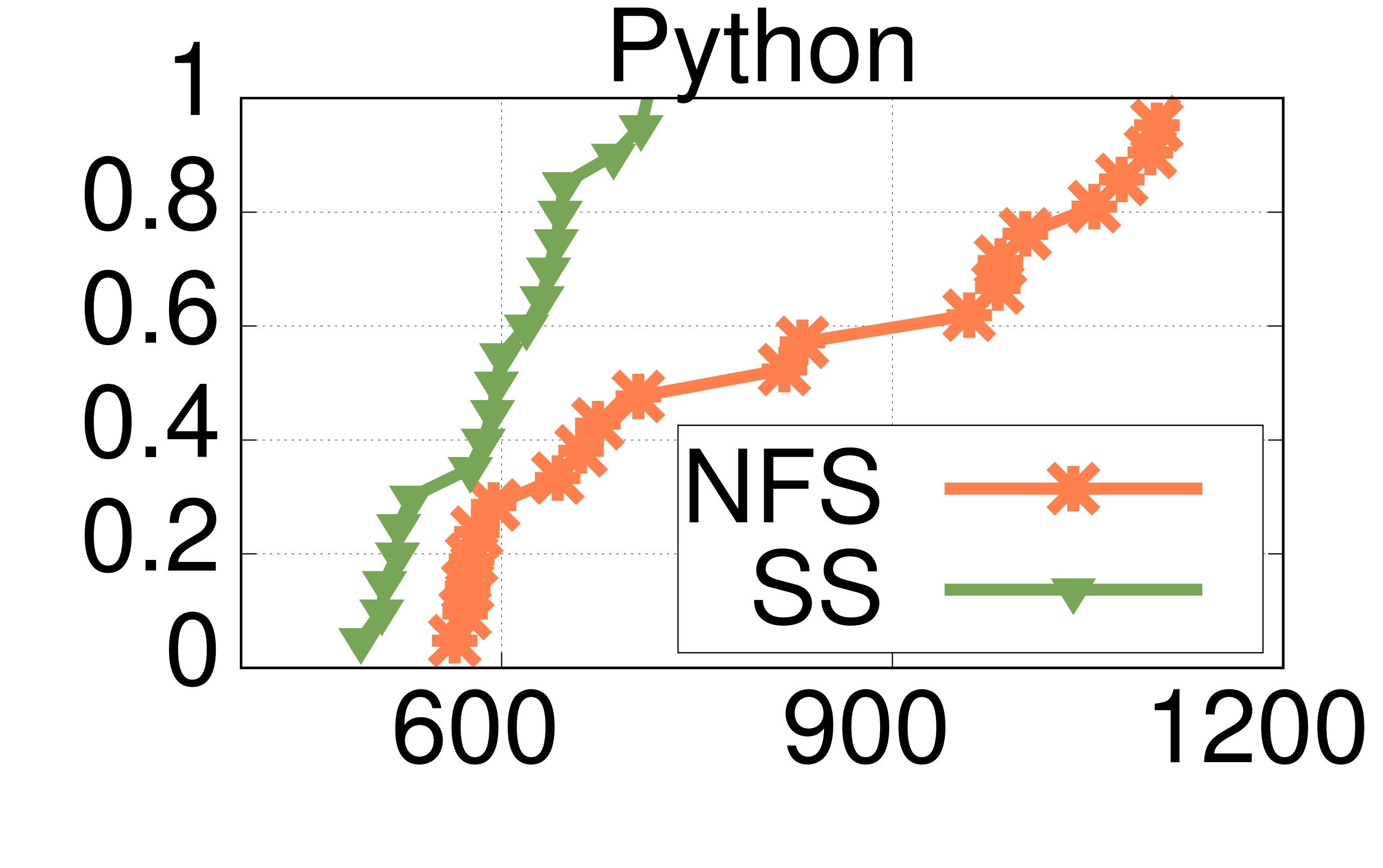}
		\end{minipage}
        \label{img-wifi-micro-python}
	\end{subfigure}
        \begin{subfigure}[t]{0.19\linewidth}
		\begin{minipage}[t]{1\linewidth}
            \includegraphics[width=1\linewidth]{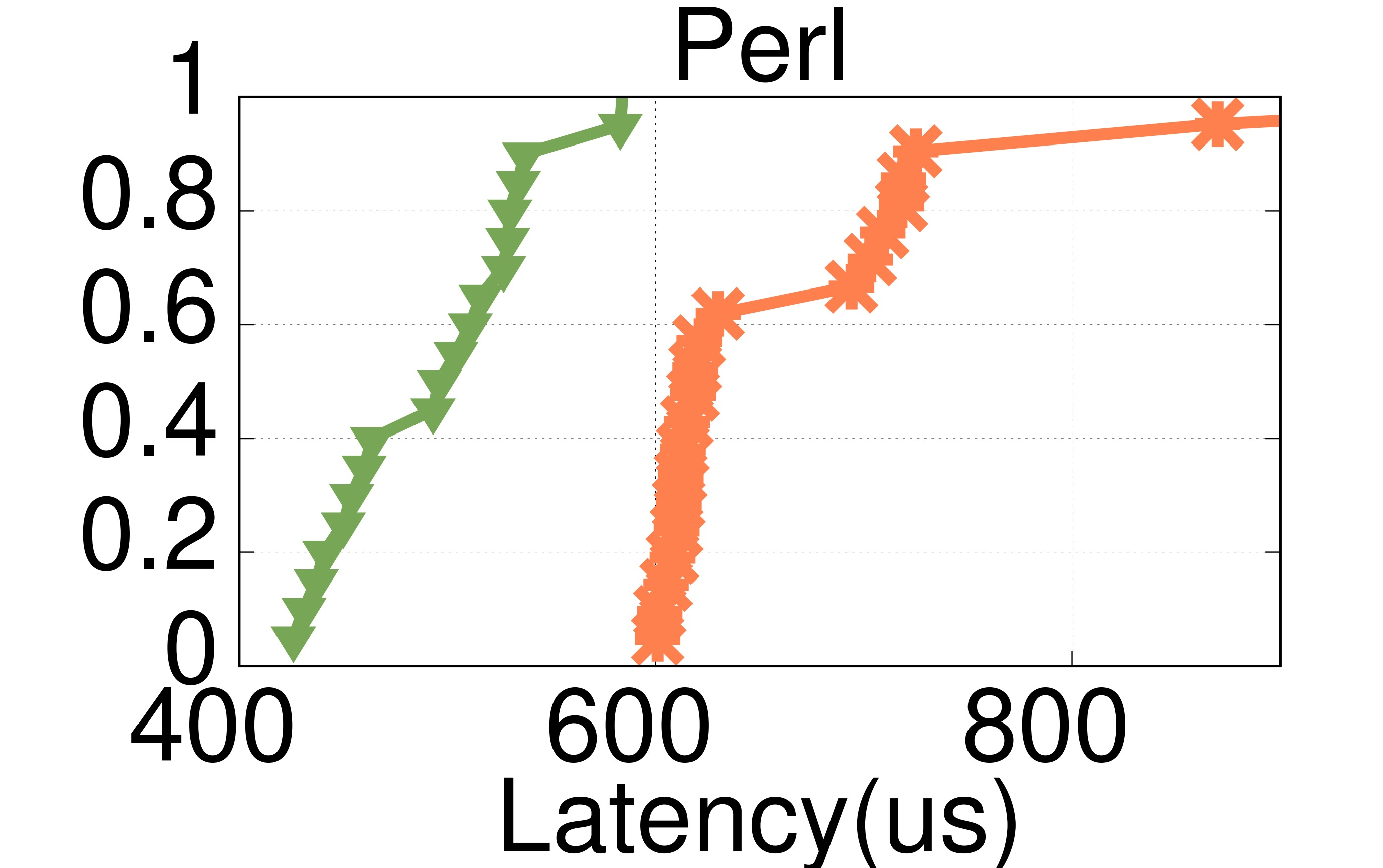}
		\end{minipage}
        \label{img-wifi-micro-perl}
	\end{subfigure}
        \begin{subfigure}[t]{0.19\linewidth}
		\begin{minipage}[t]{1\linewidth}
            \includegraphics[width=1\linewidth]{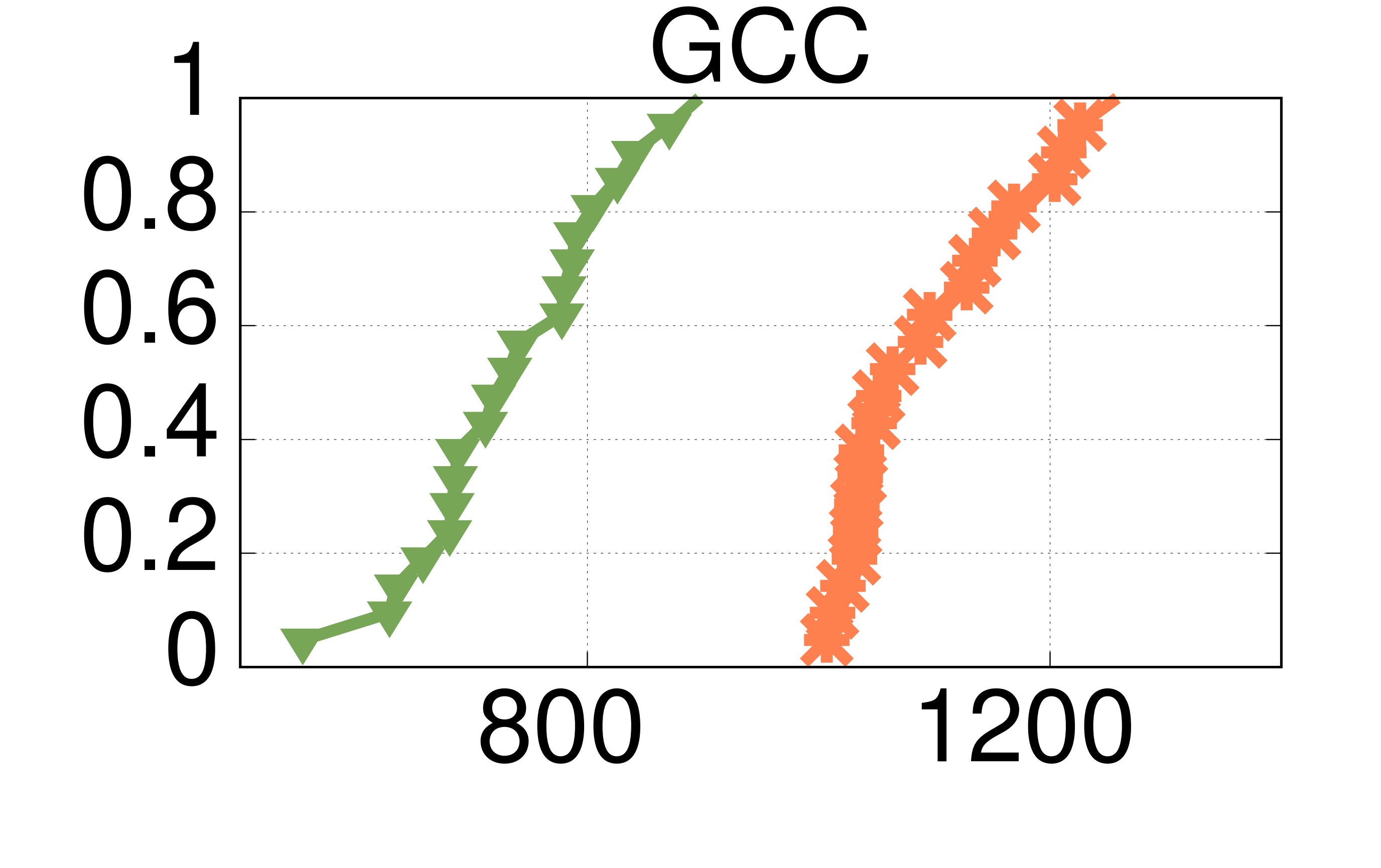}
		\end{minipage}
        \label{img-wifi-micro-gcc}
	\end{subfigure}
        \begin{subfigure}[t]{0.19\linewidth}
		\begin{minipage}[t]{1\linewidth}
            \includegraphics[width=1\linewidth]{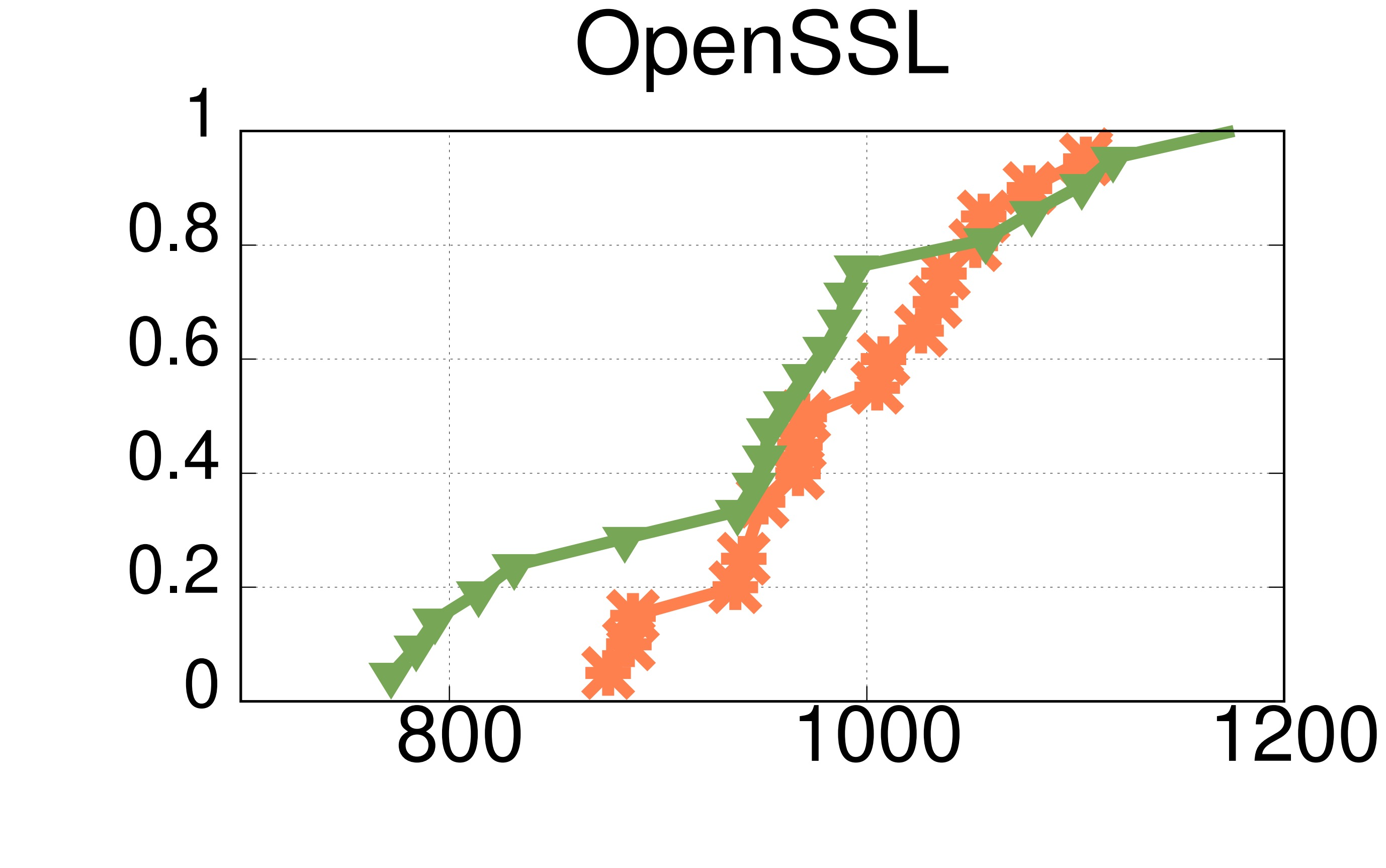}
		\end{minipage}
        \label{img-wifi-micro-openssl}
	\end{subfigure}
	\vspace{-0.5cm}
        \caption{CDF graph of the latency of fetching a single page in WiFi}
        \vspace{-0.5cm}
        \label{img-wifi-micro}    
\end{figure*}
\begin{figure*}[!htb]
	\centering
        \begin{subfigure}[t]{0.19\linewidth}
		\begin{minipage}[t]{1\linewidth}
			\includegraphics[width=1\linewidth]{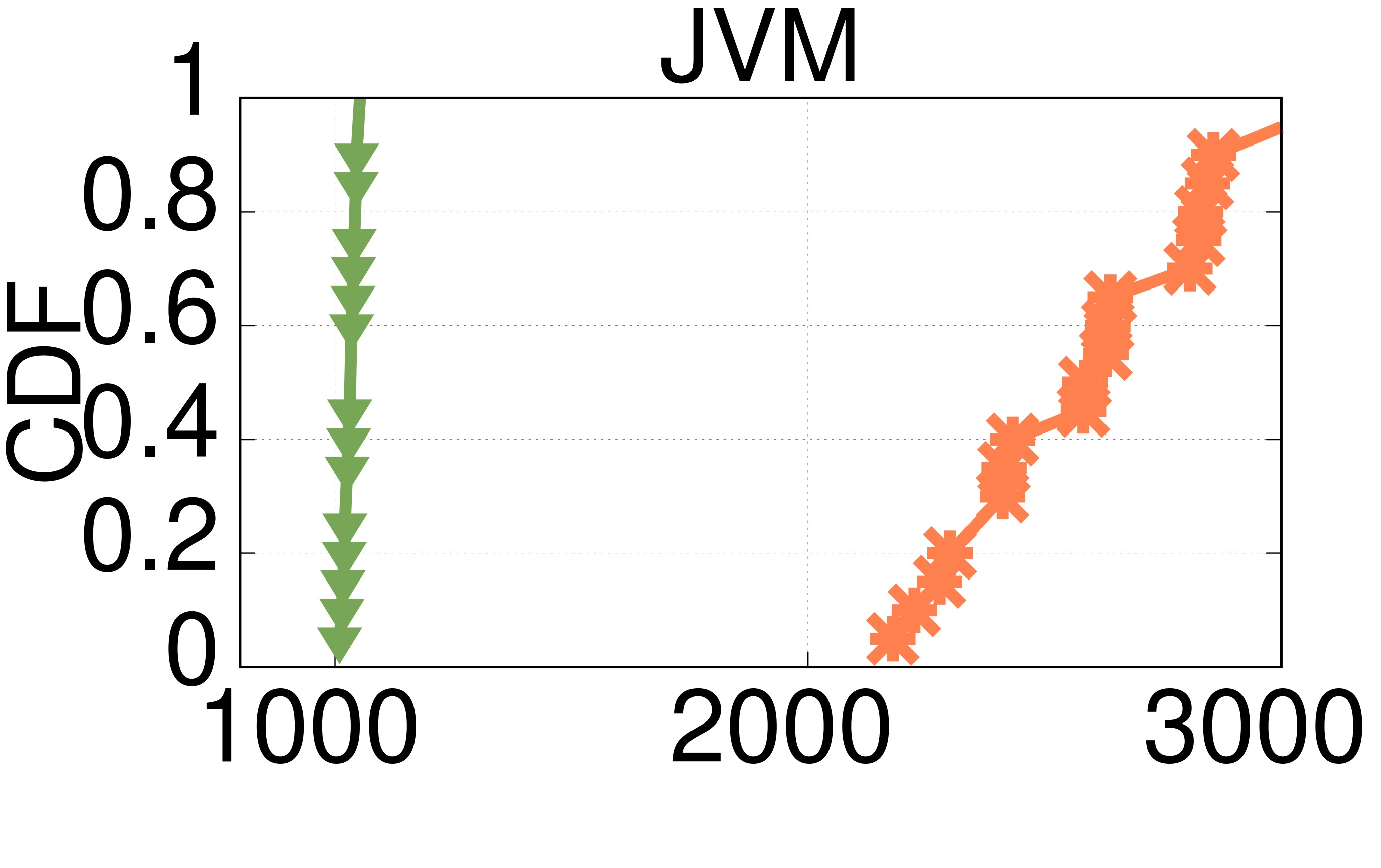}
		\end{minipage}
        \label{img-wifi-macro-jvm}
        \end{subfigure}
	\begin{subfigure}[t]{0.19\linewidth}
		\begin{minipage}[t]{1\linewidth}
			\includegraphics[width=1\linewidth]{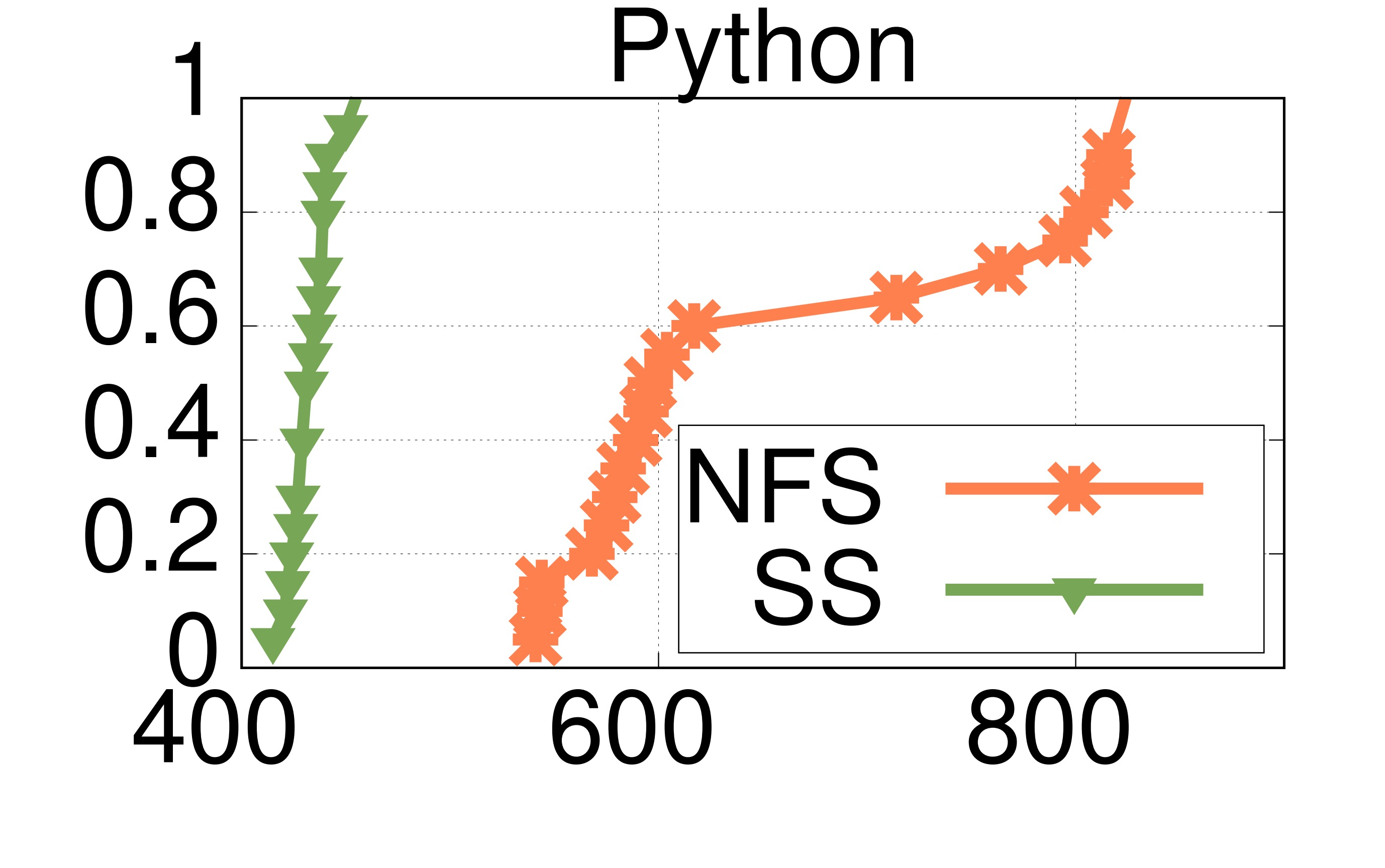}
		\end{minipage}
        \label{img-wifi-macro-python}
	\end{subfigure}
        \begin{subfigure}[t]{0.19\linewidth}
		\begin{minipage}[t]{1\linewidth}
            \includegraphics[width=1\linewidth]{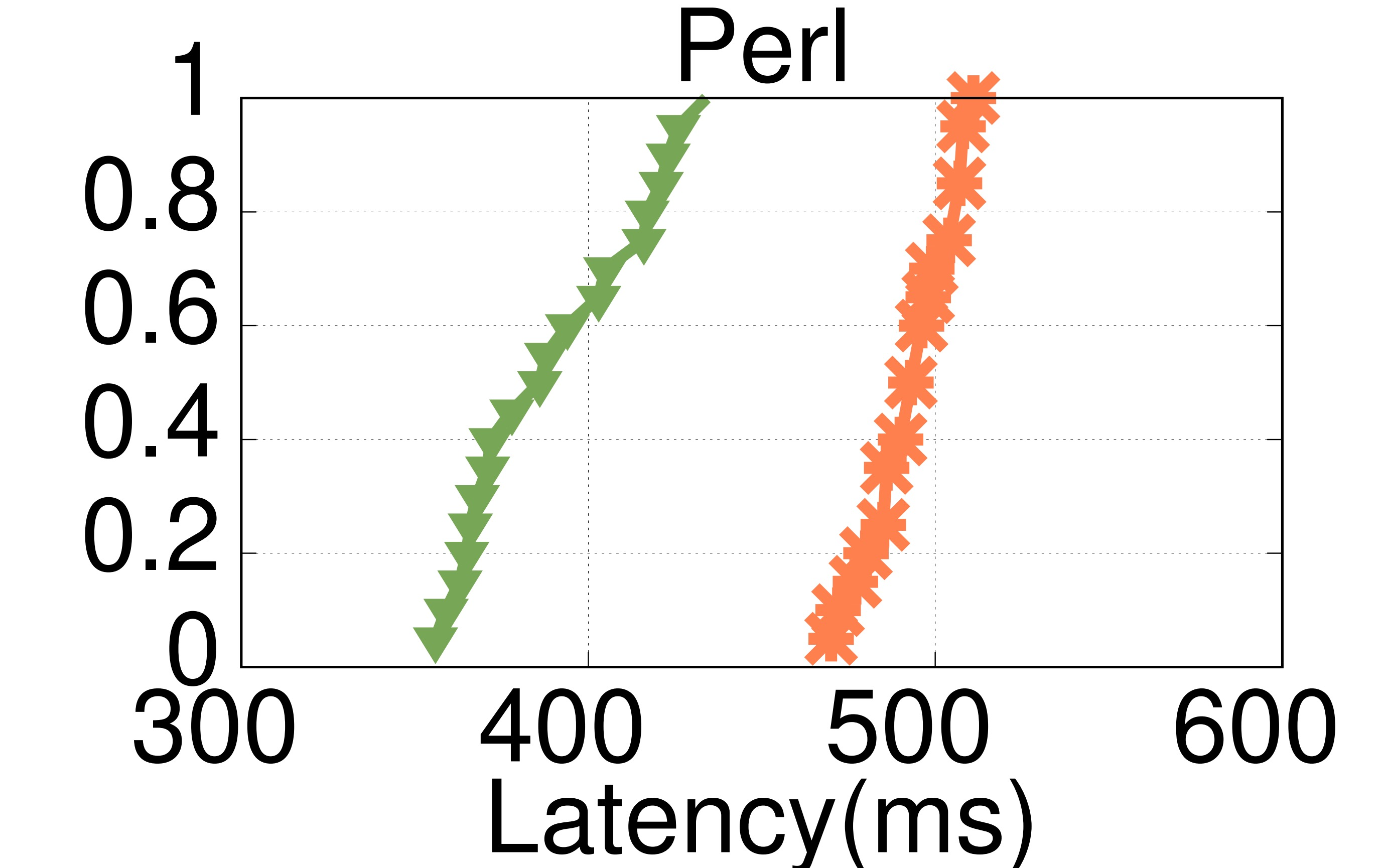}
		\end{minipage}
        \label{img-wifi-macro-perl}
	\end{subfigure}
        \begin{subfigure}[t]{0.19\linewidth}
		\begin{minipage}[t]{1\linewidth}
            \includegraphics[width=1\linewidth]{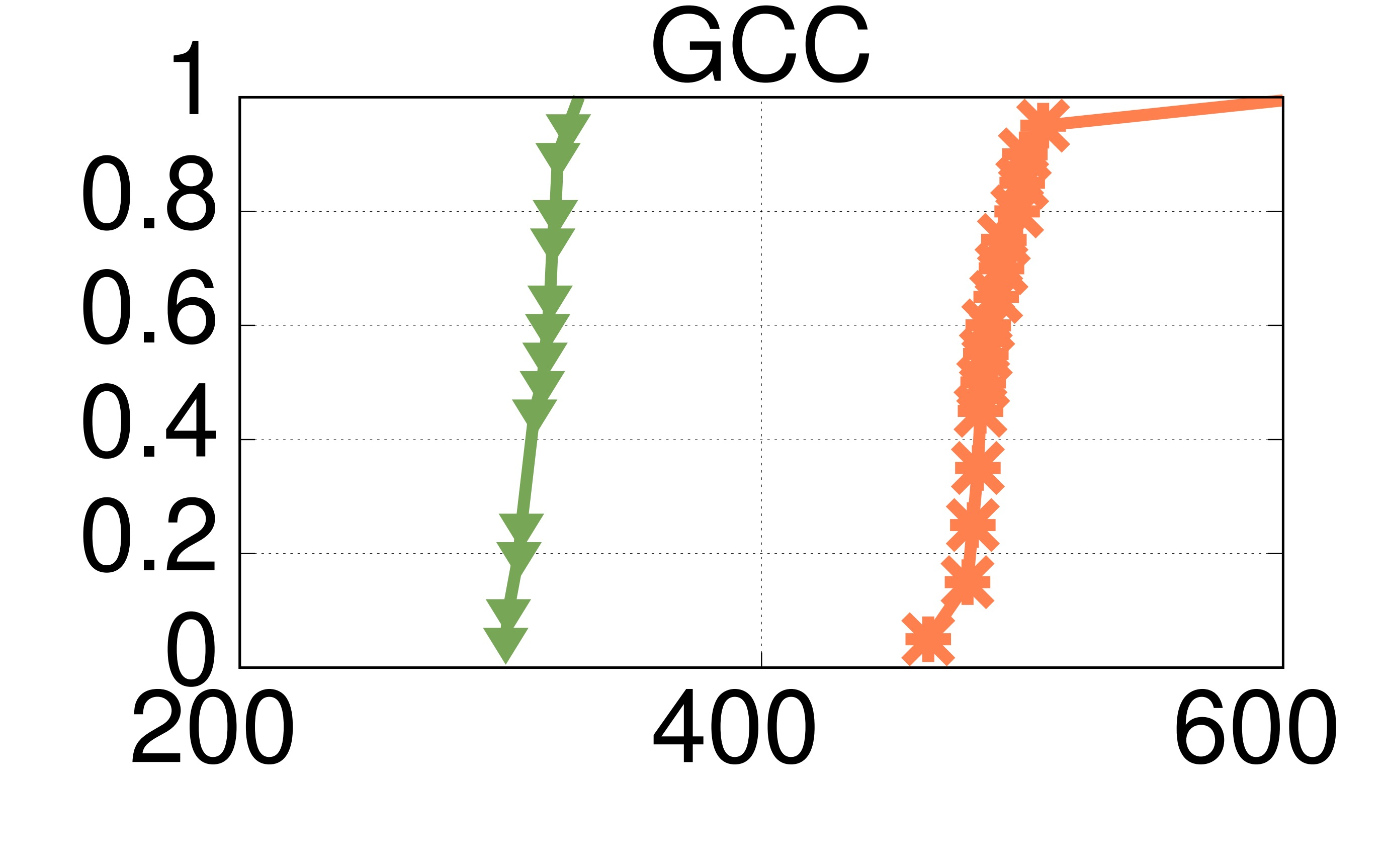}
		\end{minipage}
        \label{img-wifi-macro-gcc}
	\end{subfigure}
        \begin{subfigure}[t]{0.19\linewidth}
		\begin{minipage}[t]{1\linewidth}
            \includegraphics[width=1\linewidth]{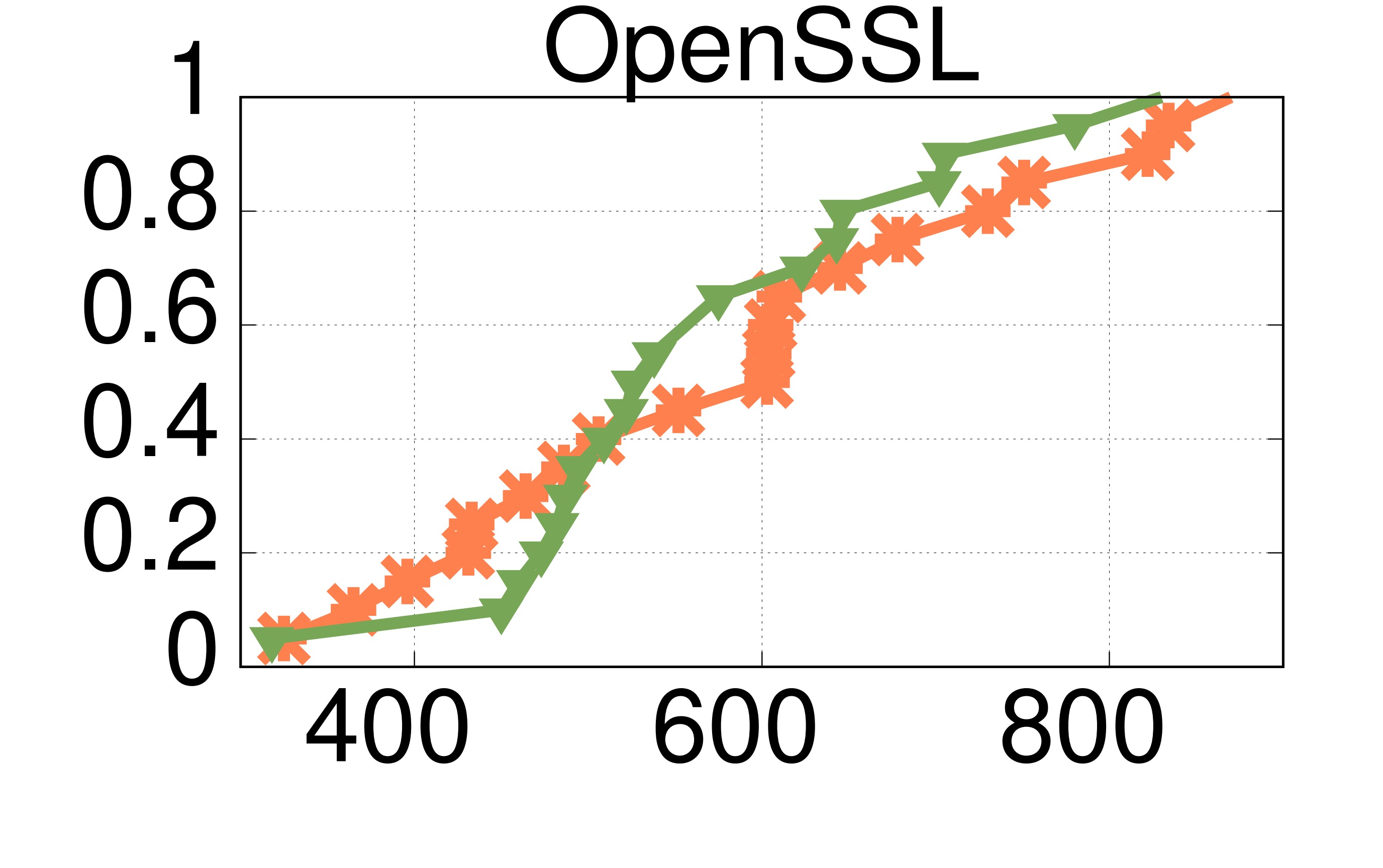}
		\end{minipage}
        \label{img-wifi-macro-openssl}
	\end{subfigure}
	\vspace{-0.6cm}
        \caption{CDF graph of applications' execution time in WiFi.}
        \label{img-wifi-macro}   
        \vspace{-0.5cm}
\end{figure*}
\subsection{WiFi}

IoT devices typically only have the ability to connect to the network through WiFi. 
We conduct an evaluation of \NAME{} in a wireless environment. 
To simulate real-world conditions, we use a household-grade wireless access point in our test.
In this evaluation, we compare NFS and \NAME{} by measuring the average time it takes for the test application to retrieve a single page in WiFi environment, as well as the overall execution time of the application. 

\textbf{\textit{Fetch a single page.}} Figure~\ref{img-wifi-micro} shows the results of our evaluation. 
From the cumulative distribution function (CDF) graph, it is evident that \NAME{} outperforms NFS, achieving a 50\% improvement. 
For applications with larger executables, such as Java, Python and Perl, \NAME{} demonstrate better performance due to its ability to make accurate segment predictions. 
However, smaller executables are less resilient to network fluctuations, leading to unstable performance.
We observe an interesting result with OpenSSL. 
The performance of OpenSSL with NFS is comparable with \NAME{}. 
This is because NFS's cache system brings an average of five network IOs, while \NAME{}'s multiple matches bring an average of seven network IOs. 
However, reducing the number of matches for the first segment can help alleviate this issue.

\textbf{\textit{Execution time.}} 
Figure \ref{img-wifi-macro} shows each application's total execution time on NFS and \NAME{}. 
The CDF graph demonstrates that \NAME{} performs more stably with fewer fluctuations than NFS, outperforming NFS by 20\% to 60.7\%. 
As mentioned in §\ref{sec-macroben}, OpenSSL's performance is significantly influenced by intensive computations and its small binary size, limiting its benefits from \NAME{}. 
Comparing Figures \ref{img-wifi-macro} and \ref{img-macrobenchmark}, \NAME{} performs 22.9\%-74.2\% worse than local Linux, while NFS performs 0.94x-2.86x worse. 
Due to GCC's IO-intensive execution logic, its performance in \NAME{} exceeds others. Native Linux suffers from a high percentage of unnecessary IO operations (Table \ref{tab-pre-read}), leading to poorer performance, while \NAME{} enhances IO efficiency, narrowing the performance gap between GCC's \NAME{} and local Linux in wireless environments. This suggests \NAME{}'s overall application execution time performance is comparable to local SD card and outperforms network-based solutions like NFS.

\begin{figure}[htbp]
    \centering
    \begin{minipage}[t]{0.48\linewidth}
        \centering
        \includegraphics[width=\linewidth]{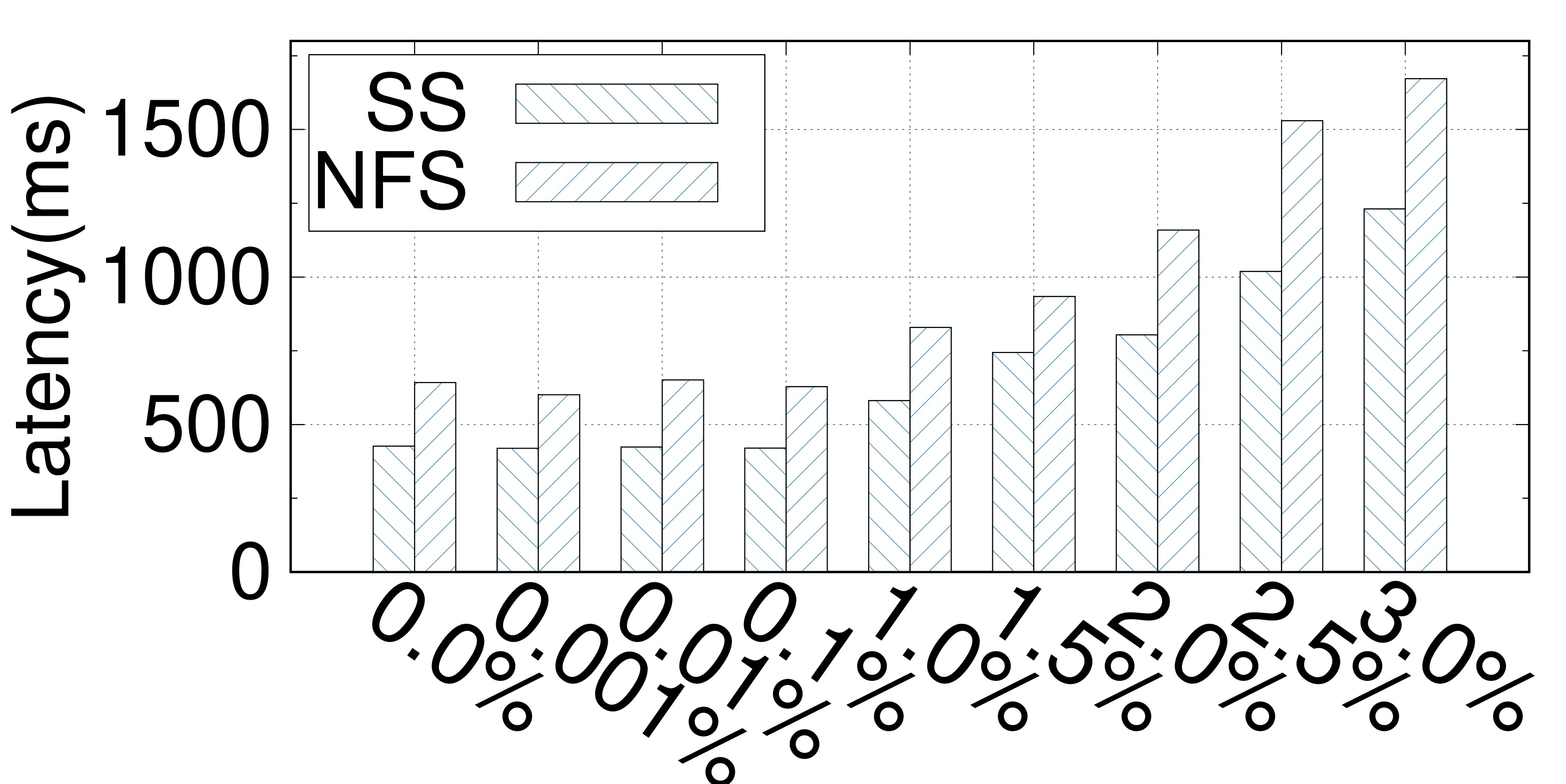}
        \caption{Latency for different packet loss rates in WiFi.} 
        \label{img-packet-loss} 
    \end{minipage}
    \begin{minipage}[t]{0.48\linewidth}
        \centering
        \includegraphics[width=\linewidth]{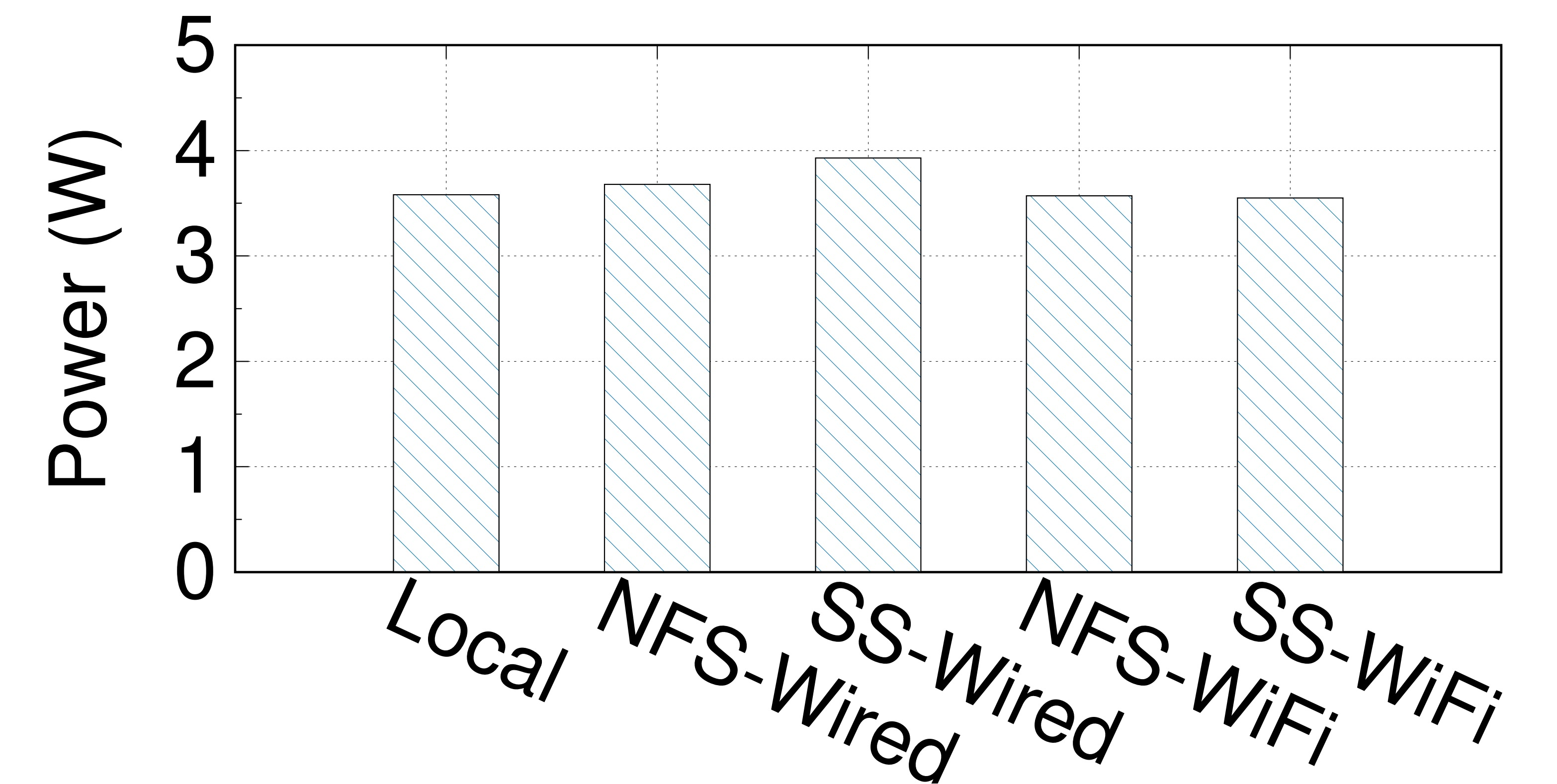}
        \caption{Power consumption for different systems. } 
        \label{img-power} 
    \end{minipage}
    \vspace{-0.5cm}
\end{figure}

\textbf{\textit{Packet Loss.}} 
Given that an IoT device's WiFi environment may experience interference, resulting in weak and unstable signals, we simulate a packet loss scenario by applying the traditional packet loss rate of WiFi and executing a Python program.
Figure~\ref{img-packet-loss} demonstrates that the performance of NFS and \NAME{} is largely unaffected when the packet loss rate is between 0.001\% and 0.1\%, indicating that these network-based solutions are effective with low loss rates. However, when the packet loss rate exceeds 1\%, both solutions show a decline in performance. It is worth noting that even under normal WiFi conditions, the performance of NFS is 50\% lower than that of \NAME{}, highlighting \NAME{}'s greater reliability and stability.    These findings suggest that, regardless of the quality of the WiFi signal, \NAME{} is a more robust and dependable solution compared to NFS.

\subsection{Power Consumption}{\label{sec-power}} 

In this study, we assess the power and energy consumption performance of IoT devices by executing Python programs on local Linux, wired NFS, wireless NFS, wired \NAME{}, and wireless \NAME{}. Figure~\ref{img-power} shows that the power consumption of \NAME{} is only 9.8\% higher than that of local Linux. Compared with the maximum of 1500 bytes for Ethernet data packets, Linux's ability to read data from local storage in 4K byte densities enables it to retrieve data with fewer interruptions, thus reducing power consumption. Moreover, the IO scheduling layer can store the blocks to be read temporarily in batches, minimizing interruptions even further. Conversely, Linux's NAPI mechanism can optimize the performance of reading network data packets by processing multiple cached Ethernet data packets in the network card through a single interrupt, thereby preventing frequent interruptions. This allows network-based solutions to achieve comparable power consumption to local storage.  Figure~\ref{img-power} shows the power consumption of \NAME{} is at worst 6.7\% higher than NFS, but it can save energy consumption due to the fact that \NAME{} can load executables with a greater network bandwidth continuously and consistently.

%% file: 6_relatedwork.tex
\section{Related work}{\label{related-work}}

\textit{\textbf{Software Debloating.}} 
Software debloating techniques include static and dynamic analysis. Among static analysis methods, early research \cite{tice2014enforcing, sun2018perses, regehr2012test, bu2013bloat} required access to the software's source code. 
Tice et al.~\cite{tice2014enforcing} used the LLVM framework and user-provided configuration information to transform the program. Quach et al.~\cite{quach2018debloating} identified and removed redundant code based on LLVM's refactoring compiler. 
 Heo et al.~\cite{heo2018effective} required users to provide usage samples to demonstrate their software usage habits. 
In order to reduce reliance on source code, Agadakos et al.~\cite{agadakos2019nibbler} proposed a static analysis technology based on binary files, and Zhang et al.~\cite{zhang2013practical} analyzed the branch instructions of the target code. 
In addition, Andriesse et al.~\cite{andriesse2016depth} obtained the call graph of all functions and removed the irrelevant code. On the other hand, dynamic analysis methods~\cite{qian2019razor, rastogi2017cimplifier, guo2011cde} recorded all the executed code during application runtime by obtaining the function call flow chart and used heuristic algorithms to expand strongly related code.

\textit{\textbf{Prefetching.}} 
Prefetching can be done at the hardware or software level to improve system performance. Hardware-level prefetching, which pre-reads data into cache or TLB, is widely studied \cite{jain2013linearizing, michaud2016best, hashemi2018learning, ayers2020classifying, jalili2022managing}. Meanwhile, software-level prefetching works similarly to hardware-level prefetching but focuses on the speed mismatch between memory and block devices \cite{fengguang2008design, readahead, al2020effectively, cao1994implementation, griffioen1994reducing, kaplan2002adaptive}. One issue with traditional file systems is that they do not allow applications to control the replacement strategy for page caches, resulting in performance degradation \cite{cao1994implementation}. To address this, a new Read-ahead framework for Linux was proposed \cite{fengguang2008design}. For emerging NVM storage, a pure software-based page management and prefetching mechanism called SPAN was introduced \cite{fedorov2017speculative}. Additionally, Leap is a prefetching solution for remote memory access that captures offset differences as access patterns and leverages majority trend-based prefetching to resist potential disturbances \cite{al2020effectively}.

%% file: 7_conclusion.tex
\section{Conclusion}{\label{conclusion}}
In this paper, we introduce \NAME{}, a lightweight executable delivery system designed for IoT devices. We propose the optimized network IO path, which redirects local disk IO to the server and enhances the performance of the network interaction between the server and clients.
We introduce an action-based block stream prefetching
mechanism by abstracting the history of access patterns as actions to reduce the page cache miss rates. We develop a normalized-variance-based method to identify and pre-read block sequences with significant differences asynchronously to avoid blocking and conceal disk IO latency.
Extensive experiments on various test benchmarks under different scenarios demonstrate the superiority of \NAME{} over the local Linux filesystem and NFS in terms of both latency and power consumption. \NAME{} performs at best 75.8\% better than Linux filesystem and 67.7\% than NFS in the wired scenario.